\documentclass[useAMS,usenatbib]{mnras}
\usepackage{natbib}
\usepackage{epsfig} 
\usepackage{scrtime}
\usepackage{graphicx}	% Including figure files
\usepackage{amsmath}	% Advanced maths commands
\usepackage{amssymb}	% Extra maths symbols
\usepackage{float}
\usepackage[caption = false]{subfig}
\usepackage{multirow}
\usepackage{rotating}
\usepackage{longtable}
\usepackage{supertabular}
\usepackage{tabularray}
\usepackage{booktabs}
\defcitealias{Rios-Lopez_et_al_2021}{Paper I}

\newcommand{\logd}{\log}

\newcommand{\mincir}{\raise-3.truept\hbox{\rlap{\hbox{$\sim$}}\raise4.truept\hbox{$<$}\ }}
\def\mean#1{\left< #1 \right>}
\title[2D Surface Brightness of Large Galaxies II]{2D Surface Brightness Modelling of Large 2MASS Galaxies II: The Role of Classical Bulges and Pseudobulges on Galaxy Scaling Relations and its implication for Supermassive Black Hole Formation}
\author[R\'ios-L\'opez,L\'opez-Cruz,A\~norve, et al. ]
{Emmanuel R\'ios-L\'opez,$^{1,2,3}$\thanks{(e-mail:riloemm@inaoep.mx)} Omar L\'opez-Cruz,$^{1}$
Christopher A\~norve,$^{4}$
\newauthor
Mabel Valerdi,$^{1,5}$ Victoria R. Dufrane,$^{6}$ 
\& Erick  A. Rodr{\'\i}guez-Hern\'andez.$^7$
\\ 
$^{1}$Instituto Nacional de Astrof\'isica, \'Optica y Electr\'onica (INAOE), Coordinación de Astrofísica, Tonantzintla, Puebla, M\'exico. \\
$^{2}$Instituto de Astrof\'isica de Canarias (IAC), V\'ia L\'actea s/n, 38205, La Laguna, Tenerife, Espa\~na.\\
$^{3}$Universidad Pedag\'ogica del Estado de Sinaloa (UPES), Castiza s/n, 80027, Culiac\'an, Sinaloa, M\'exico.\\
$^{4}$Facultad de Ciencias de la Tierra y el Espacio (FACITE), Universidad Aut\'onoma de Sinaloa (UAS), Culiac\'an, Sinaloa, M\'exico.\\
$^{5}$Consejo Nacional de Humanidades, Ciencias y Tecnologías (CONAHCyT), Av. Insurgentes Sur 1582, 03940, Ciudad de México, México. \\
$^{6}$Facultad de F{\'\i}sica, Universidad Veracruzana (UV), Xalapa, Veracruz, M\'exico.\\
$^{7}$Facultad de Instrumentación  Electr\'onica, Universidad Veracruzana (UV), Xalapa, Veracruz,M\'exico.}
\date{Accepted XXX. Received YYY; in original form ZZZ}

\pubyear{2024}

\begin{document}
\label{firstpage}
\pagerange{\pageref{firstpage}--\pageref{lastpage}}
\maketitle

\begin{abstract}
We have generated 2D-multicomponent surface brightness (SB) modelling for 100 galaxies in the Large Galaxy Atlas (LGA) together with 19 nearby cD galaxies using the near-infrared (NIR) images from  2MASS ($J, H {\, \rm and}\, K_s$). Our final sample of 119 galaxies includes cD galaxies, Virgo cluster galaxies, group galaxies, and field galaxies. We revisited known scaling relations (SRs) involving structural parameters, as well as those involving supermassive black holes (SMBHs) and ultramassive black holes (UMBHs). Refining the SRs, we also revisited the bulge classification and considered the Fundamental Plane (FP) and its projections, as well as other SRs, such as the colour-magnitude relation (CMR), Tully-Fisher relation (TFR) and luminosity concentration relation (LCR). Classical bulges follow the same relations as elliptical galaxies, while pseudobulges are usually outliers. The NIR colours of classical bulges and pseudobulges indicate that their ages are not radically different despite their spread in luminosity, but we noticed that classical bulges are more luminous than pseudobulges, therefore, this property provides a complementary bulge classification criterion. We included pseudobulges from other studies to strengthen the tendencies seen for pseudobulges in our sample. From the SRs for BHs, we found that pseudobulges do not follow SRs for early-type galaxies and classical bulges. Additionally, the lack of correlation between BHs and discs may indicate these structures have not coevolved. From the revision of SRs, we present a sample of galaxies likely to host SMBHs or UMBHs, which are suitable for dynamical BH mass determination from the ground.

\end{abstract}

\begin{keywords}
(galaxies:) quasars: supermassive black holes - galaxies: bulges - galaxies: discs - galaxies: structure - galaxies: elliptical and lenticular, cD
\end{keywords}

\section{Introduction}
\label{Intro}
The properties of galaxies show mutual correspondences known as scaling relations (SRs). Indeed, SRs have been found among galaxy parameters such as mass, colour, size (e.g., effective radius $r_e$), luminosity ($L$), surface brightness (SB), kinematics (velocity dispersion, $\sigma$); rotation velocity ($V_{rot}$), and star formation, among others \citep[see][hereafter DMC, for a recent comprehensive review]{2021FrASS...8..157D}. SRs help us to explore galaxy formation or find critical stages during their evolution. However, some aspects (such as gas accretion, galaxy mergers, and star formation, among others) of the origin of SRs have remained unexplained \citep[e.g.,][]{Peng_2007, 2015ARA&A..53...51S}. Cosmological initial conditions may be more relevant than previously thought \citep[e.g.,][]{2020MNRAS.498.4386P}. Hence, SRs provide insight into the formation and evolution of galaxies.

Some SRs can be understood, at least heuristically, by assuming Virial equilibrium and accepting a form for the mass-to-light ratio ($M/L$) \citep[e.g.,][DMC]{Kormendy-Djorgovski_1989,2011ARA&A..49..301V}. The space of parameters resulting after galaxy evolution shows regularities, suggesting shared processes \citep{2008Natur.455.1082D} that might regulate internal structure formation and black holes (BH) coevolution. Hereafter, we recognise three types, namely of BHs: supermassive BH (SMBH: $10^{6}\, \mathrm{M_{\sun}} \leq M_\bullet  < 10^{10}\, \mathrm{M_{\sun}}$), ultramassive BH \citep[UMBH: $10^{10} \leq M_\bullet < 10^{11}\, \mathrm{M_{\sun}}$,][]{2012MNRAS.424..224H}, and the theoretical stupendously large BH \citep[SLAB: ${\rm M_{\bullet}}\gtrsim 10^{11}\; \mathrm{M_{\sun}}$, ][]{2021MNRAS.501.2029C}. Some processes associated with the evolution of galaxies are secular, while others are violent, such as major mergers. In that regard, secular processes are inhered to the galaxy or driven by the environment, e.g., ram pressure stripping, galaxy harassment and minor mergers \citep[e.g.,][]{Kormendy-Kennicutt_2004,2013seg..book....1K}. 

The coevolution of galaxies and SMBHs \citep[see][for a review]{Kormendy-Ho_2013} is suggested by the tight SRs found between the SMBH mass and luminosity \citep{Kormendy-Richstone_1995, Marconi-Hunt_2003}, $\sigma$ \citep{Ferrarese-Merritt_2000, Gebhardt_et_al_2000, Ferrarese-Ford_2005} and stellar mass of the bulge \citep{Magorrian_et_al_1998}. Other physical parameters seem to correlate with SMBH,  for example, with the S\'ersic index \citep{Graham_2001, Graham_2007}, the virial mass of the host galaxy \citep{Ferrarese_et_al_2006}, the cusp radius ($r_{\gamma}$) in the most massive elliptical galaxies \citep{Lauer_et_al_2007, Lopez-Cruz_et_al_2014}. The most luminous cD galaxies might follow different SRs, suggesting environment-driven formation mechanisms, primarily found in clusters of galaxy  \citep[e.g.,][]{Kormendy-Ho_2013,Lopez-Cruz_et_al_2014,Mehrgan_et_al_2019}.

 The Faber-Jackson relation \citep[FJR,][]{Faber-Jackson_1976} indicates a correspondence between luminosity and stellar $\sigma$ ($L \propto \sigma^4$) among early-type galaxies (ETGs), i.e. E+S0. The FJR can also be stated in terms of $M/L$. \citet{Kormendy_1977} found a correlation between size and SB ($I_e\propto r_e$); in this case, the SB is measured at the effective radius $r_e$. This is known as the Kormendy Relation (KR). Later, the FJR and the KR were recognised as projections of a more general SR called the Fundamental Plane (FP). The FP relates kinematics, SB and scale size \citep[][]{Djorgovski-Davis_1987, Dressler_et_al_1987}:  $I_e \propto \sigma^{\alpha}r_e^{\beta}$, where $I_e$ is the SB at $r_e$. Considering a homologous system and assuming the Virial Theorem; then, $\alpha=2$ and $\beta=1$ \citep[e.g.,][]{Faber_et_al_1987}, this is called {\em the virial scaling}. We have that luminosity and size depend on distance, but kinematics does not; hence, these properties combined in the FP give suitable distance indicators \citep[e.g.,][]{Dressler_et_al_1987, Djorgovski-Davis_1987}. However, the observed FPs deviated from the virial scaling; this was called the ``tilt'' of the FP \citep[e.g.,][]{Faber_et_al_1987, Pahre_et_al_1998, Scodeggio_et_al_1998, Bernardi_et_al_2003}. The solution was envisaged by  \citet{Faber_et_al_1987}; nevertheless, the tilt problem remained open for many years. As it was found, $M/L$ variations, indeed, produce the tilt of the FP,  in agreement with the prediction of \citeauthor{Faber_et_al_1987} \citep[see reviews by][also DMC]{2014RvMP...86...47C,Cappellari_et_al_2013, Cappellari_2015}. 

Closely related to the FJR is the Tully-Fisher relation \citep[TFR,][]{Tully-Fisher_1977}, which shows a connection between galaxy luminosity and the line width of the 21 cm line integrated over the whole galaxy, i.e., $L\propto V_{rot}^{\gamma}$ where $\gamma\simeq3$ in the optical, 
$\gamma\simeq4$ in the NIR \citep[e.g.,][]{2001ApJ...550..212B}. This SR was found earlier by \citet{1969AJ.....74..859R}; nevertheless, \citet{Tully-Fisher_1977} popularised its use as a distance indicator. The TFR depends on the morphological type and the photometric band used to measure the galaxy's luminosity \citep[e.g.,][]{2011ARA&A..49..301V,2001ApJ...563..694V}. The TFR has been generated for large galaxy surveys recently
\citep[e.g.,][]{Courteau_et_al_2007, 2007ApJ...668...94H,2013IAUS..289..296G,2013AJ....146...86T}.

Another important SR is the Luminosity-Size Relation \citep[LSR; e.g.,][]{1996ApJ...460L..89B,1996ApJ...464L..63S,1996ApJ...465L.103S,2002ASSL..274..269L, Nair_et_al_2010}, which describes that brighter galaxies are progressively larger. Galaxies in dense environments populate the LSR at $z \la 1.5$ \citep[e.g.,][]{1997ApJ...477L..17S}. Likewise, the Colour-Magnitude Relation (CMR) has been widely studied using large samples of cluster ETG galaxies. \citep[e.g.,][]{Lopez-Cruz_et_al_2004,2013pss6.book..265B} and, similarly, to the LSR, the CMR has been found already in place at $z \la 1.5$ \citep[e.g.,][]{1998ApJ...501..571G,2006ApJ...644..759M}. For colours generated by combining ultraviolet and optical bands or optical bands with NIR bands, the more luminous galaxies are redder on the CMR \citep[e.g.,][]{2016AJ....152..214S}; this has been explained by assuming that the stellar populations in more massive galaxies are more metal-rich \citep[e.g.,][and references therein]{2006MNRAS.370.1106G}. This paper extends the early works by \citet{1959PASP...71..106B} and \citet{1978ApJ...223..707S} by including contributions of classical bulges, pseudobulges and discs separately to the CMR.  

\begin{table*}
%	\tiny
	\centering
	\caption{Subsample of morphologically selected cD galaxies ordered, as the Abell cluster catalogue, by right ascension.}
	\label{Table_cDs}
	\tabcolsep 6 pt
	\begin{tabular}{llccccl}
	\hline
~~Name & Abell cluster & Distance & $K_s$ &$M_{K_s}$& $\log\, L_{K_s} $ & ~~~$\sigma$ \\
\multicolumn{1}{c}{} & & [Mpc] & [mag.]& [mag.]& [$L_{\odot,K_s}$] & [km/s]  \\
~~~~(1) & ~~~~~(2) & (3) &  (4) & (5) & (6)& ~~(7) \\
\hline
UGC~428 & A0077-BCG & 305.9 & $10.65 \pm 0.04$ &$-26.66 \pm 0.08$ & 11.97 $\pm$ 0.01 &$-$ \\%& 12.02 $\pm$ 0.02\\
Holm~15A & A0085-BCG & 247.0  & $09.92 \pm 0.02 $& $-26.96 \pm 0.07$ & 12.09  $\pm$ 0.03 &$310 \pm 15$ \\
UGC~716 & A0150-BCG & 258.0 & $10.44 \pm 0.06$ & $-26.52 \pm 0.09$ & 11.92  $\pm$ 0.02 & $245 \pm 35$ \\
MCG+00-05-040 & A0261-BCG & 203.8 & $10.93 \pm 0.07$ & $-25.55 \pm 0.10$ & 11.53  $\pm$ 0.04 & $408 \pm 33$ \\
UGC~2438 & A0399-BCG & 320.9 & $10.09 \pm 0.05$ &$-27.37 \pm 0.08$ &  12.26 $\pm$ 0.02 & $230 \pm 30$ \\
UGC~2450 & A0401-BCG & 329.7 & $10.03 \pm 0.04$ & $-27.49 \pm 0.08$ & 12.30  $\pm$ 0.02 &$367 \pm 35$ \\
MCG-02-12-039 & A0496-BCG & 144.6 & $08.19 \pm 0.10$ &$-27.60 \pm 0.12$  & 12.35 $\pm$ 0.05 & $241 \pm 14$ \\
PGC~025714 & A0754-BCG & 247.2 & $10.76 \pm 0.04$ & $-26.16 \pm 0.24$ & 11.77  $\pm$ 0.02 &$350.8 \pm 16.9$ \\
PGC~025777 & A0754-${\rm BCG_2}$ & 247.2 & $10.98 \pm 0.09$ & $-25.91 \pm 0.11$ & 11.67 $\pm$ 0.04 & $274.8 \pm 20.3$ \\
NGC~2832 & A0779-BCG & 103.0 & $08.10 \pm 0.04$ & $-26.93 \pm 0.08$ & 12.08  $\pm$ 0.02 & $325.8 \pm 10.8$ \\
UGC~5515 & A0957-BCG & 204.8 & $08.75 \pm 0.13$ & $-27.74 \pm 0.15$ & 12.40 $\pm$ 0.05 & 299.5 $\pm$ 22.9 \\
NGC~3551 & A1177-BCG & 143.5 & $08.42 \pm 0.07$ &$ -27.31 \pm 0.10$ & 12.23 $\pm$ 0.03 & $255 \pm 17$ \\
NGC~4839 & A1656-${\rm BCG_3}$& 103.2 & $08.73 \pm 0.03$ &$-26.30 \pm 0.07$ & 11.83 $\pm$ 0.01 & $278.6 \pm 4.3$ \\
NGC~4874 &A1656-${\rm BCG_2}$& 103.2 & $08.30 \pm 0.02$ &$-26.73 \pm 0.07$ & 12.00  $\pm$ 0.01 & $271.9 \pm 4.3$ \\
NGC~4889 & A1656-BCG & 103.2 & $08.05 \pm 0.10$ & $-26.98 \pm 0.07$ & 12.10 $\pm$ 0.01 & $393 \pm 5.3$ \\
NGC~5778 & A1991-BCG & 262.8 & $10.36 \pm 0.09$ & $-26.64 \pm 0.11$ & 11.96 $\pm$ 0.04 & $279 \pm 25$ \\
IC~1101 & A2029-BCG & 360.0 & $09.57 \pm 0.05$ &$-28.09 \pm 0.08$ &12.54 $\pm$ 0.02 & $359 \pm 12$ \\
NGC~6166 & A2199-BCG & 134.5 & $08.68 \pm 0.03$ &$-26.91 \pm 0.07$ & 12.07  $\pm$ 0.01 & $281 \pm 39$ \\
\hline
\multicolumn{6}{l}{}\\
\multicolumn{6}{l}{\textbf{Notes.} (1) cD galaxy name. (2) Alternative name, ${\rm BCG_n}$ is the nth BCG in cluster-cluster merger}\\
\multicolumn{6}{l}{mergers. (3) Luminosity distance computed with the redshift as corrected to the CMB rest}\\
\multicolumn{6}{l}{ frame from NED. (4) Total apparent magnitude in the $K_s$ band from this work. (5) Stellar}\\
\multicolumn{6}{l}{velocity dispersions ($\sigma$) taken from \texttt{Hyperleda} \citep{Paturel_et_al_2003}. (6) Logarithmic $K_s$}\\
\multicolumn{6}{l}{band luminosity in solar units.}\\
\end{tabular}
\end{table*}

\begin{table*}
%	\tiny
	\centering
	\caption{Galaxies with reported SMBH masses or SMBH mass upper limits, ordered by increasing BH mass.}
	\label{Table_SMBH}
	\tabcolsep 6 pt
	\begin{tabular}{llllccc}
	\hline
Name & Type& Distance & $\log M_{\bullet}$ & Bulge Type & Method & Ref. \\
\multicolumn{2}{c}{}& [Mpc] & [${\rm M}_{\odot}$] & & & \\
(1) &(2) & (3) & (4) &  (5) & (6) & (7)\\
\hline
NGC~5102&SA$0^-$ & 3.2 & $5.94 \pm 0.38$ & P & S & 2 \\
NGC~4945 &Scd& 3.58 & $6.15 \pm 0.30$ & C & M & 2 \\
NGC~4826 &Sab& 7.27 & $6.19 \pm 0.11$ & C & S & 1 \\
M~32 &E2& 0.80 & $6.39 \pm 0.19$ & C & S & 9 \\
M~61 &SABbc& 12.3 & $6.58 \pm 0.17$ & P & G & 2 \\
NGC~1365 &SB(s)b& 17.8 & $6.60 \pm 0.30$ & C & G & 2 \\
M~96 &Sab& 10.62 & $6.88 \pm 0.08$ & P & S & 4 \\
NGC~253 &Sc& 3.5 & $7.00 \pm 0.30$ & P & G & 2 \\
M~91 &SBb& 17.9 & $7.25 \pm 0.29$ & C & G & 3 \\
M~88 &Sbc& 16.5 & $7.30 \pm 0.08$ & C & G & 4 \\
NGC~3953 &SBb& 15.4 & $7.33 \pm 0.29$ & P & G & 3 \\
M~106 &SABbc& 7.27 & $7.58 \pm 0.03$ & C & M & 1 \\
NGC~613 &SB(rs)b& 15.4 & $7.60 \pm 0.35$ & P & G & 3 \\
NGC~1672 &SB(s)b& 11.4 & $7.70 \pm 0.10$ & P & G & 2 \\
NGC~7582 &SB(rs)ab& 22.3 & $7.74 \pm 0.20$ & C & G & 1 \\
M~81 &Sb& 3.60 & $7.81 \pm 0.13$ & C & S & 1 \\
M~31	&Sb& 0.77 & $8.16 \pm 0.16$ & C & S & 12 \\
NGC~1316 &S0& 20.95 & $8.23 \pm 0.07$ & C & S & 11 \\
NGC~3377 &E5-6& 10.99 & $8.25 \pm 0.25$ & C & S & 10 \\
NGC~5005 &SAB(rs)bc& 14.6 & $8.27 \pm 0.23$ & C & G & 3 \\
NGC~4697 &E6& 12.54 & $8.31 \pm 0.10$ & C & S & 10 \\
NGC~1097 &SBb& 24.9 & $8.38 \pm 0.04$ & C & G & 2 \\
NGC~4636 &E0-1& 13.7 & $8.58 \pm 0.22$ & P & G & 3 \\
M~105 &E1& 10.70 & $8.62 \pm 0.11$ & C & S & 9 \\
NGC~4526 &SAB(s)$0^0$:& 16.44 & $8.65 \pm 0.12$ & C & G & 1 \\
%NGC~4593 & 38.5 & 6.86 $\pm$ 0.21 & C & G & 3 \\
M~104 &SA(s)a edge-on& 9.87 & $8.82 \pm 0.04$ & C & S & 1 \\
M~63 &Sb& 8.9 & $8.94 \pm 0.10$ & C & G & 2 \\
NGC~3115 &S$0^-$ edge-on& 9.54 & $8.95 \pm 0.10$ & C & S & 1 \\
M~49 &E2& 16.72 & $9.40 \pm 0.05$ & C & S & 13 \\
M~60	&E2& 16.46 & $9.67 \pm 0.10$ & C & S & 14 \\
M~87 &cD pec& 16.80 & $9.81 \pm 0.05$ & C & Imaging & * \\
NGC~4889 &cD4& 103.2 & $10.32 \pm 1.6$ & C & S & 17 \\
Holm~15A &cD& 247.0 & $10.60 \pm 0.8$ & C & S & 16 \\
\multicolumn{6}{c}{Upper limits on $M_{\bullet}$}\\
M~33  &Sc& 0.82 & 3.19 & P & S & 15 \\
M~110 &E5 pec& 0.82 & 4.38 & P & S & 8 \\
NGC~6503 &Sc& 5.3 & 6.30 & P & S & 6 \\
M51a &Sc& 7.9 & 6.32 & P & G & 5 \\
M~101 &SAB(rs)cd& 7.0 & 6.41 & P & S & 6 \\
IC~342 &Sc& 3.73 & 6.50 & P & 5 & 5 \\
NGC~3351 &SBb& 9.3 & 6.78 & P & G & 5 \\
M~100 &Sbc& 14.2 & 6.84 & P & G & 5 \\
M~85 &SA$0^+$(s) pec& 17.88 & 7.11 & C & & 7 \\
NGC~2903 &SAB(rs)bc& 10.4 & 7.34 & P & G & 5 \\
NGC~3675 &Sb& 12.4 & 7.56 & C & G & 5 \\
NGC~4579 &SAB(rs)b& 23.0 & 8.32 & C & G & 5 \\
\hline
\multicolumn{7}{l}{}\\
\multicolumn{7}{l}{\textbf{Notes.} (1) Galaxy name. (2) Galaxy type is taken from NED. (3) Distance taken from the same  }\\
\multicolumn{7}{l}{reference as the SMBH mass. (4) SMBH mass or upper limit. }\\
\multicolumn{7}{l}{(5) Our bulge classification: classical bulge (C) or pseudobulge (P) (see \citetalias{Rios-Lopez_et_al_2021}). (6) Method used}\\
\multicolumn{7}{l}{to measure SMBH mass: S, G and M for stellar, gas and maser dynamics, respectively.}\\
\multicolumn{7}{l}{(7) References for SMBH masses:}\\
\multicolumn{7}{l}{1: \cite{Kormendy-Ho_2013},}\\
\multicolumn{7}{l}{2: \cite{Sahu_et_al_2019_2}, 3: \cite{van_den_Bosch_2016}, }\\
\multicolumn{7}{l}{4: \cite{deNicola_et_al_2019}, 5: \cite{Beifiori_et_al_2012}}\\
\multicolumn{7}{l}{6: \cite{Kormendy_et_al_2010}, 7: \cite{Gultekin_et_al_2011},}\\
\multicolumn{7}{l}{8: \cite{Valluri_et_al_2005}, 9: \cite{vd_Bosch-d_Zeeuw_2010}, }\\
\multicolumn{7}{l}{10: \cite{Schulze-Gebhardt_2011}, 11: \cite{Nowak_et_al_2008},}\\
\multicolumn{7}{l}{12: \cite{Bender_et_al_2005}, 13: \cite{Rusli_et_al_2013},}\\
\multicolumn{7}{l}{14: \cite{Shen-Gebhardt_2010}, 15: \cite{Gebhardt_et_al_2001},}\\
\multicolumn{7}{l}{16: \cite{Mehrgan_et_al_2019}, 17: \cite{McConnell_et_al_2012},}\\
\multicolumn{7}{l}{\textbf{*} from direct imaging \cite{EHT_2019_6}.} \\
%\multicolumn{6}{l}{\cite{EHT_2019_6} from direct imaging}\\
%\multicolumn{7}{l}{}\\
\end{tabular}
\end{table*}

This is a companion paper to \citet{Rios-Lopez_et_al_2021}  \citepalias[hereafter][]{Rios-Lopez_et_al_2021} presenting a broad study of galaxy SRs using morphological parameters generated from 2D photometric decomposition for a sample of 101 2MASS bright galaxies shown in \citetalias{Rios-Lopez_et_al_2021}. Additionally, we consider 18 nearby cD galaxies to explore the high end of the SRs for SMBHs and UMBHs. Kinematic data and BH masses were taken from the literature. Since most sources in our sample are local luminous galaxies, 2MASS telescopes' angular resolution can, in detail, solve internal structures, such as bulges, discs, bars, and arms (see \S \ref{sb_phot}). The advantage of near-infrared (NIR) observations is that they penetrate gas and dust, catching the light from low-mass and old stars; these are the bulk of the stellar population. This is a major baryonic component in most galaxies \citep[e.g.,][]{1996ASSL..209...65F, 1996A&AS..118..557D,1996A&A...313..377D,Jarrett_2000,Jarrett_et_al_2000b,Jarrett_et_al_2003}.

This paper is organised as follows: \S\ref{Data-Met} presents the data analysed in this paper, which was obtained mainly from \citetalias{Rios-Lopez_et_al_2021}, along with measurements of SMBH masses compiled from the literature and photometry performed here to the cD galaxies. A summary of the photometric decomposition methodology is also presented. In \S\ref{Res} and \S\ref{Dis}, we present and discuss our results for galaxy SRs, as well as those including SMBH masses; also, results are presented in the context of bulge types (namely, classical bulges and pseudobulges). Finally,  in \S\ref{Con}, we present our conclusions. Unless indicated otherwise,  as in \citetalias{Rios-Lopez_et_al_2021}, we have adopted the following cosmology: $\mathrm{H_0}=70\,h_{70}\;\mathrm{km\,s^{-1}\, Mpc^{-1}}$,$\:\Omega_{\Lambda}=0.7\:\mathrm{and}\: \Omega_m=0.3$, throughout this paper.

\section{Galaxy Sample and Methodology}\label{Data-Met}

\citetalias{Rios-Lopez_et_al_2021} details sample selection, observations, and photometric decompositions used in this paper. We briefly describe the sample selection and methods employed to generate SB measurements with {\texttt{GALFIT}}, error analysis, corrections due to galactic extinction and bulge classification strategies. Below, we also outline how we selected and analysed a group of 18 cD galaxies added in this paper.
 
\subsection{The Sample} \label{Sample}
 
We worked on the Large Galaxy Atlas \citep[LGA,][]{Jarrett_et_al_2003}\footnote{LGA data available at \url{https://irsa.ipac.caltech.edu/applications/2MASS/LGA/atlas.html}} to draw the 101 brightest nearby galaxies ($z\leq 0.01$, $K_s\leq 9.37$ mag) out of the 551 LGA galaxies (see Table 1 in \citetalias{Rios-Lopez_et_al_2021}). Our sample includes galaxies from the Virgo cluster galaxies, groups, cD galaxies, and field galaxies, but 68 \% are spiral galaxies \citep[cf.,][]{Jarrett_2004}. The observations have enough sensitivity and angular resolution to account for the main galaxy components, such as bulges, discs, bars, and spiral arms \citep[for more details, see][and \S \ref{sb_phot} in this paper]{Jarrett_2000,Jarrett_et_al_2003}. Galaxies' distances were taken from the literature, mainly based on the mean from determinations listed in NED (primary distance indicators were taken if available), while for most of the galaxies with SMBH masses, the distance was taken from the same reference and for cD galaxies we adopted the luminosity distance generated from the cluster's redshift corrected for the CMB. 

The SB photometry of late-type galaxies (LTG) was done using LGA mosaics, but for ETGs we used 2MASS IRSA image tiles\footnote{2MASS image tiles service \url{https://irsa.ipac.caltech.edu/applications/2MASS/IM/interactive.html##pos}}. The referee of \citetalias{Rios-Lopez_et_al_2021} pointed out that LGA's mosaics suffered from sky oversubtraction and provided us with relevant information\footnote{See the extended discussion on this issue in \url{https://wise2.ipac.caltech.edu/staff/jarrett/2mass/ellipticals.html}} \citep[see also][]{Schombert_2011}. The problem arose, as we understood, for not allowing large enough apertures to account for the SB extension of ETG galaxies. We worked with both sets of images, finding that using LGA mosaics, the integrated total magnitudes for ETGs were underestimated by $\sim 0.35$ mag \citepalias[see \S 3.6 and Fig. 3 in][]{Rios-Lopez_et_al_2021}. This result agrees with  \citet{Schombert_2011, 2016AJ....152..214S} and the more sensitive NIR observations in  \citet{2024MNRAS.527..249Q}; yet, for the SB photometry of LTGs, we did not resort to the image tiles: we stayed with the LGA mosaics because they had no distinguishable sky over-subtraction as shown by  Fig. 3 in \citetalias{Rios-Lopez_et_al_2021}, which shows the comparison of the SB profiles of the LTG M61. 
We refer the reader to \citetalias{Rios-Lopez_et_al_2021} for further details on sample selection and photometry generation.

\subsubsection{cD Galaxy Selection} \label{cD}
 
Complementing \citetalias{Rios-Lopez_et_al_2021}, we have included the photometric decomposition of 18 nearby cD galaxies in the $K_s$ band of 2MASS using IRSA image tiles for the reasons given above (\S \ref{Sample}). Table \ref{Table_cDs} lists the cD galaxies, including galaxy name, cluster name, distance in Mpc, total magnitudes in the $K_s$ band, luminosity in terms of $L_{\sun}$ in the $K_s$ band and velocity dispersion data. Since only two UMBHs have been measured directly, our aim in this section is to extend the SRs for SMBHs at the high-mass end and find UMBHs and SLABs candidates for future studies.  

\citet{1965ApJ...142.1364M} recognised cD galaxies as supergiant galaxies, reaching up to 4 magnitudes brighter than the characteristic magnitude $M^*$; hence, they are the most luminous galaxies \citep[e.g.,][]{1987IAUS..127...89T}. However, cD galaxies were recognised initially as counterparts of powerful radio sources by \citet{1964ApJ...140...35M}. cD galaxies are, in most cases, the BCGs, but two or three cD's can be present in some clusters. When this occurs, we usually deal with cluster-cluster mergers (e.g., Coma has three cD galaxies: NGC~4839, NGC~4874 and NGC~4889).

\citet{1987ApJS...64..643S} introduced a new classification scheme for cD galaxies based on the presence of a faint light halo with shallow-slope SB profile extending below $\mu_V=24 \; \mathrm{mag \, arcsec^2}$. This feature is barely picked by direct inspection of photographic plates. Hence, this definition is difficult to apply, as the detection of low-SB features can be affected by telescope optics, pixel size and sky brightness. Moreover, the detection of low-SB structures is complicated as $z$ increases due to cosmological SB-dimming, $\mu_0\sim\mu_e(1+z)^{-4}$, where $\mu_0$ and $\mu_e$ are the observed brightness and the emitted SB, respectively. Substantial discrepancies arose when  CCD observations did not recover the cD extended low-SB halos reported earlier using photographic plates \citep[e.g., A1413-BCG, NGC~ 6166;][respectively]{2002ApJ...575..779F, 2015ApJ...807...56B}, due to these problems, \citet{1987ApJS...64..643S}'s classification scheme has been avoided and have been called brightest cluster galaxy (BCG). BCGs are tacitly considered a homogeneous class \citep[e.g.,][]{2007MNRAS.379..867V}. Nevertheless, we suggest that this view doesn't hold, in general. Indeed, the BCGs are mostly cD galaxies in rich clusters of galaxies; however, in poorly populated irregular ones, the BCG could be a spiral, a lenticular (S0) galaxy, or a giant elliptical galaxy \citep[cf.,][]{2019MNRAS.482.4084L}; therefore, BCGs are a mixed class, at best. cD galaxies, on the other hand, can be found either in rich or poor clusters, as well as in local overdensities \citep[e.g.,][]{1975ApJ...199..545M,1977ApJ...211..309A, 2002ApJ...575..779F}; however, the presence of more than one cD, radio relics,  and radio haloes are direct indicators of cluster-cluster mergers \citep[e.g.,][]{2002ASSL..272....1S,2008LNP...740....1S,2008LNP...740..143F,2024A&A...686A..55L}.

We suggest that cD galaxies conform to a homogenous class.  We propose that instead of the shape of the SB profile, cD galaxies should be identified by the luminosity contrast between the BCG and satellite galaxies \citep[][original definition]{1965ApJ...142.1364M}. It happens that this is the same defining property in the classification scheme introduced by \citet[BM,][]{1970ApJ...162L.149B} for clusters as a whole. These cluster types are quantifiable by using variations to the strategy proposed by \citet{1978ApJ...222...23D}; nevertheless, cluster classification schemes based on the degree of regularity are hardly quantifiable. We focus on the BM type I clusters with an easily distinguishable BCG dominant over the brightness of the satellite galaxies. Hence, following \citet{1993AJ....106..831H}, we call cD galaxies those BCGs in BM I or BM I-II (intermediate class) clusters, which at the same time were classified according to \citet[RS,][]{1971PASP...83..313R} as a cD cluster. Our classification scheme is complementary to \citet{2012MNRAS.427.2047T}, where BCGs in BM I clusters, exclusively, were called cD galaxies. This way, the extended low-SB halo becomes a secondary property, as suggested by \citet{1997ApJ...475L..97L}. See \citet[][and references therein]{2003gafe.conf..109L} for a short review of the philosophy and applications of morphological classifications in Astronomy. 

We took BM and RS types from the compilations of \citet{1977ApJS...34..381L} and \citet{1987ApJS...63..555S}, respectively. We also checked the classifications directly by inspecting  DPOSS, 2MASS or SDSS images. Complementary to our selection, we included the cD galaxies embedded in rich substructures.  For example, in the Coma Cluster, we selected NGC~4839 (A1656-$\mathrm{BGC{_3}}$, the third brightest cluster galaxy), NGC~4874 (A1656-$\mathrm{BGC{_2}}$) and NGC~4889 (A1656-BCG), as well as in Abell 754 Cluster, we selected two cD galaxies PGC~025714 (A754-$\mathrm{BGC{_2}}$) and PGC~025777 (A754-BCG). These rich clusters are known cluster-cluster mergers \citep[e.g.,][]{2008LNP...740..143F,2011A&A...536A...8P}. The bright cD galaxies in Table \ref{Table_cDs} have apparent total magnitudes at least two magnitudes brighter than the 2MASS detection limit of $K_s=13.1$ for extended sources \citep{Jarrett_et_al_2000a}. The full sample is given in Table \ref{Table_cDs}, where we have introduced a subindex to the galaxy's name to indicate the second or the third BCG. The cD galaxy M~87  was already included in \citetalias{Rios-Lopez_et_al_2021}; therefore, we have considered 19 cD galaxies in total in this paper. 

\subsubsection{The Final Sample}

The final sample considered in this paper contains 119 nearby high-SB galaxies drawn from representative structures in the cosmic web. Although we tried to cover the entire Hubble sequence, 68 \%  of the galaxies in the sample are spirals, including intermediate and barred galaxies; hence, ETGs are underrepresented 
\citepalias[see Table 1 in ][]{Rios-Lopez_et_al_2021}. However, our sample is neither complete by volume nor by magnitude.

\subsection{SMBH Masses} \label{SMBH_Masses_sample}

We found the SMBH masses for 31 galaxies in common with \citet{Rios-Lopez_et_al_2021} sample; we present them in Table \ref{Table_SMBH}.  The reported masses were measured directly using stars or gas dynamics, the kinematics of astrophysical masers inside accretion discs, or the modelling of the BH shadow. Their values and methods used to measure the SMBH masses are listed in columns (4) and  (6), respectively, as well as the corresponding reference in column (7). The morphological classifications are in column (2), which were taken from NED (see  \citet{2013seg..book..155B} for a comprehensive review on galaxy types); in column (5), we provide bulge/pseudobulge classification taken from \citetalias{Rios-Lopez_et_al_2021}. Upper limits were found for 12 sources; although the upper limits of the SMBH mass were not included in the regression analyses, they are depicted on the plots just for completeness.

From all the cD galaxies selected in this paper, only M87, Holm~15A, and NGC~4889 have dynamically measured BH masses, while for the cD galaxy M~87, we adopted the SMBH mass reported by \citet{EHT_2019_6}. Thus, in our results reported below for the SMBH scaling relations, we also include the BH masses of Holm~15A, NGC~4889 and M~87.

For completeness, we have included additional pseudobulges only in the $K_s$ band; such data were taken from \citet{Kormendy-Ho_2013} and \citet{deNicola_et_al_2019} and references therein. We must point out that photometric parameters for pseudobulges for some references were derived using one-dimensional methods, unlike our 2D methodology. The results for correlations with pseudobulges from the literature are labelled with an asterisk (*)  in Table \ref{Table_results_fits_M-rels}. An inspection of Table \ref{Table_SMBH} shows that, as indicated by the galaxy classification, the earliest galaxies, those with more prominent classical bulges, host more massive SMBH. We noticed below that bulges are more luminous than pseudobulges, representing an additional criterion for bulge/pseudobulge classification.  

\subsection{Velocity Dispersions ($\sigma$)}

Data on the central stellar velocity dispersionn $\sigma$ (column 8 in Table 1 from \citetalias{Rios-Lopez_et_al_2021}) were taken from 
\citet{Ho_et_al_2009}, as well as from \texttt{HyperLeda} database\footnote{\url{http://leda.univ-lyon1.fr/leda/param/vdis.html}} \citep{Paturel_et_al_2003}. Data for effective stellar velocity dispersion within one effective radius $\sigma_e$ were taken from the studies that reported SMBH dynamical-mass measurements. At the same time, H$\alpha$ or HI rotation velocity ($V_{rot}$) data were also taken from \texttt{HyperLeda}.

\subsection{2D decomposition with {\texttt{GALFIT}}}

We apply {\texttt{GALFIT}}\footnote{\url{https://users.obs.carnegiescience.edu/peng/work/GALFIT/GALFIT.html}} \citep{Peng_et_al_2002, Peng_et_al_2010} to our sample, a popular 2D decomposition algorithm to extract the structural parameters for each of the modelled components: magnitudes, effective radii, S\'ersic index, among others. {\texttt{GALFIT}} models the SB distribution using parametric functions and can fit many sources simultaneously. In addition, {\texttt{GALFIT}} uses most of the pixels on a galaxy image, allowing for more degrees of freedom than 1D schemes \citep[e.g.,][]{Schombert_2012}, which considers the simultaneous modelling of subcomponents individually, accounting for variations in the internal geometrical distribution separately. {\texttt{GALFIT}} is started using input parameters generated by SExtractor \citep{Bertin-Arnouts_1996}. We have generated our packages to manipulate {\texttt{GALFIT}} output; in this paper, we use {\texttt{EllipSect}}\footnote{ {\texttt{EllipSect}} can be downloaded from \url{https://github.com/canorve/EllipSect/tree/v2.2.4}}\citep{Ellipsect_2025}, which is a Python tool designed to take {\texttt{GALFIT}} output to generate SB profiles, as well as to extract and compute absolute magnitude, luminosity, effective radii, Akaike information \citep[AIC;][]{akaike74} and Bayesian information criteria \citep[BIC;][]{schwarz78} used to distinguish among different {\texttt{GALFIT}} models for galaxies, among other parameters. As we indicated above, a complete discussion on SB modelling can be found in 
\citetalias{Rios-Lopez_et_al_2021}.

For the cD galaxies in Table \ref{Table_cDs}, we applied the same methodology as in \citetalias{Rios-Lopez_et_al_2021}. A single  S\'ersic component was usually enough to model the 2MASS data. The sum of two S\'ersic models was used when necessary. For the cD galaxy in A85, Holm~15A, we reported a total apparent magnitude of  $K_s=10.08\pm 0.03$ using growth curve analysis \citep{Lopez-Cruz_et_al_2014}. The new total apparent magnitude found in this paper is slightly brighter, $K_s=9.92\pm 0.02$ (see Table \ref{Table_cDs}) after taking into account the effects of the PSF and integrating to infinity the S\'ersic fit using {\texttt{GALFIT}}. Considering the photometric errors and that only a few of the bright neighbouring galaxies were masked by  \citet{Lopez-Cruz_et_al_2014}, we suggest that our estimates for the total luminosity of Holm~15A  are consistent with each other. We compared total magnitudes estimated by growth-curve analysis of cluster galaxies and total magnitudes measured after modelling the SB considering seeing effects and integrating the profiles to infinity; we found that the former total magnitudes are  $\sim0.1\, \%$ fainter, with a dispersion of $0.4\;\%$ than the latter ones (L\'opez-Cruz et al., 2024, in preparation). 

\subsubsection{Model Selection, uncertainties, and magnitude corrections}

We considered the following three-way strategy to model  galaxies' SB: 

\begin{enumerate}
    \item {\bf single} component represented by a S\'ersic profile \citep{Sersic_1968} for elliptical galaxies;
    \item {\bf bulge+disc} components for disc galaxies, where the bulge and disc are modelled by S\'ersic and exponential functions, respectively;
\item {\bf bulge+disc+bar} components for barred galaxies, where a S\'ersic profile adjusted the bar. 
\end{enumerate}
The exponential profile \citep{Freeman_1970} is a special case of the S\'ersic function when $n$ = 1, while the bar structure can be modelled with $n$ $\leq$ 0.5 \citep[e.g.,][]{Greene_et_al_2008, Peng_et_al_2010}. For galaxies hosting active galactic nuclei (AGN), a point spread function (PSF) profile is used as an extra component to account for the contribution of the unresolved nuclear structure associated with the AGN. If AGNs or bars are not accounted for in the modelling, the S\'ersic index is unreliable or may increase to unrealistic values of $n >$ 10 \citep[e.g.,][and references therein]{2015ApJS..219....4S}. In some cases, we considered an additional S\'ersic component \citepalias[see][]{Rios-Lopez_et_al_2021} to account for substructures that resembled extended stellar envelopes or embedded discs \citep[e.g.,][]{Lasker_et_al_2014}. Also, referring to the spheroidal component parameters through this paper could be either the bulge or the central S\'ersic component of E galaxies modelled with two components. Our strategy guarantees the accurate modelling of bulge parameters since we use a multi-component approach to account for the main galaxy substructures, namely disc, bulge, bar and AGN \citep[e.g.,][]{2005MNRAS.362.1319L,2015ApJS..219....4S}. 

We have developed an alternative approach to derive more realistic uncertainties since the ones provided by {\texttt{GALFIT}} are underestimated. Again, we refer the reader to Appendix A of \citetalias{Rios-Lopez_et_al_2021}.

\subsubsection{Absolute Magnitudes and Luminosities} 

We calculated the absolute magnitudes and luminosities for each band. Below, we show a sample calculation for the $K_s$ band: 
\begin{equation} \label{AM}
M_{K_s}= K_s-DM-k_{corr}-A_{K_s},
\end{equation}
where $K_s$ is the apparent magnitude in the $K_s$ filter; the distance modulus is given by $DM= 5\log D_L+25$, where $D_L$ is the luminosity distance in Mpc, for cD galaxies, we used the cluster's redshift corrected for the CMB; $k_{corr}$ is the $k$-correction in the $K_s$ band, generated using the calculator by \citet{Chilingarian_et_al_2010}\footnote{See the online service \url{http://kcor.sai.msu.ru}}, using the galaxy's redshift and either $J$-$K_s$ or $H$-$K_s$ colours as input, for the cD galaxies we set $J$-$K_s=1.0$ mag, the mean value for bulges and ETGs (see Fig. \ref{CMR}). Alternatively,  a  k-correction approximation given by  $k^{\prime}_{corr}=-5.6\log(1+z)$ is found as a  fit to  \citet{2001MNRAS.326..745M} data for  $z\leq 0.3$ galaxies. This simple approximation agrees with \citeauthor{Chilingarian_et_al_2010}; $A_{Ks}$ is the galactic extinction according to  \citet{Schlegel_et_al_1998} in the direction of the target galaxy, generated with the aid of the calculator provided by the NASA/IPAC Infrared Science Archive\footnote{Galactic Dust Reddening and Extinction \url{https://irsa.ipac.caltech.edu/applications/DUST/}}.  

Luminosities were calculated by assuming that the absolute magnitude of the Sun in the $K_s$-band is $M_{(\odot,K_s)}=3.27 \,\rm{mag}$ \citep[e.g.,][]{2018ApJS..236...47W}; hence, 
\begin{equation}\label{lum}
L_{K_s}= 10^{[(M_{(\odot,K_s)}-M_{K_s})/2.5]}\;L_{(\odot, K_s)}, 
\end{equation}
where $M_{K_s} $is given by Equation \ref{AM}. Table \ref{Table_cDs}  
include absolute magnitudes and luminosities in the $K_s$ band generated using Equations \ref{AM} and \ref{lum}. The same procedure was followed for the $H$ and $J$ bands.

\subsection{Robust Linear Regressions} \label{Linear_regressions}

We fitted the SRs in this paper using the \texttt{LtsFit}\footnote{\url{https://pypi.org/project/ltsfit/}} routine implemented by  \citet{Cappellari_et_al_2013}. This is an outlier-resistant fitting scheme based on the Least Trimmed Squares \citep[LTS,][]{1987rrod.book.....L} minimization. It also allows for uncertainties in both axes and provides a measurement of the intrinsic scatter of the linear regression.

Thus, we express our results using the following linear form:
\begin{equation} \label{line_fit}
y =  m\left(x - \langle x \rangle \right) + b  + \epsilon_y, \\
\end{equation}
\noindent that results after a $\chi^2$ minimization to a distribution on $N$ points
$(x_i,y_i)$, where $m$ and $b$ are the slope and zero-point of the linear fit, respectively;  the reference value or pivot $\langle x \rangle$ is the mean of $x_i$, which is introduced to reduce the uncertainty in $b$, and 
$\epsilon_y$ is the intrinsic scatter around the dependent variable, i.e., $y$.  \texttt{LtsFit} proceeds iteratively rejecting outliers until the reduced $\chi^2_\nu=1$, i.e., the $\chi^2$ per degree of freedom $\nu=N-2$. The $1\sigma$ errors in the slope ($\sigma_m$) and zero point ($\sigma_b$) are generated from the covariance matrix. In contrast, the error in the dispersion ($\sigma_{\epsilon_y}$) is computed by incrementing $\epsilon_y$ until $\chi=\nu-\sqrt{2\nu}$ \citep[see \S 3.2 in][for further details]{Cappellari_et_al_2013}. 

This way of fitting a straight line given by Equation \ref{line_fit} to the data points simplifies (e.g., $\langle y \rangle =b$) and eases the propagation of errors. For extrapolations of any value of $x$ the error in $y$ is given by \citep[e.g.,][]{1989snpp.book.....L,Berendsen_2011}:
\begin{equation}\label{properr}
    \sigma_{y}^2= \sigma_m^2\left(x - \langle  x \rangle \right)^2+ \sigma_b^2.
\end{equation}
We have generated the $1\sigma$ confidence intervals by solving Equation \ref{properr} for each SR fit; grey-shaded regions indicate them, while the scatter $\epsilon_y$ is indicated by dashed lines in the figures below.

Equation \ref{line_fit} may seem unfamiliar; nevertheless, we can easily compare with published results by identifying $m$ as the slope and 
\begin{equation}\label{intercept}
b_\circ \equiv b - m\langle x \rangle,
\end{equation}

\noindent
as the intercept. This simple transformation allows us to recover the familiar $y=mx+b_{\circ}$. 

 We should remark that \texttt{LtsFit} is set for data whose errors are Gaussian and uncorrelated; if the errors are correlated, the code must be modified accordingly \citep[see][and references therin]{Cappellari_et_al_2013}.
  
\citet{Cappellari_et_al_2013} applied sophisticated Bayesian analysis and confirmed that \texttt{LtsFit} is robust to outliers. They also found insignificant differences using either approach despite their fundamental conceptual differences. We corroborated \citeauthor{Cappellari_et_al_2013} claim: we applied a Bayesian regression approach to our data to compare with the fits given in Table \ref{Table_results_fits_FP_projections}. There is plenty of software available for Bayesian two-variable linear regression, however, only a few consider errors in both variables. For example, consider the FJR in the $J$ band for the entire sample (\S\ref{FJR} and Table \ref{Table_results_fits_FP_projections}). After applying \texttt{linmix}, a Bayesian routine developed by \citet{Kelly_2007} that considers errors in both variables, we get $m_B=-0.112\pm0.008$ and $b_B=-0.31\pm0.17$, the errors reported in this Bayesian approach are derived from the posterior distribution. In contrast, using \texttt{LtsFit} we found $m=-0.11\pm0.01$ and $b_{\circ}=-0.27\pm 0.21$. Hence, the results are indeed indistinguishable as the fitted parameters are the same within the errors; besides, for this example, the errors coming from either a covariance matrix analysis or a posterior distribution are comparable. We found excellent agreement when we checked for other SRs in Table \ref{Table_results_fits_FP_projections}, so we kept and chose to report the  \texttt{LtsFit} fits.

To facilitate comparisons with previous studies, we report fitted parameters using the following format: $m\pm \sigma_m$, $b\pm\sigma_b$, $b_{\circ} \pm \sigma_{b_{\circ}}$, 
$\epsilon_y\pm\sigma_{\epsilon_y}$, and $\langle x \rangle$ in Tables \ref{Table_results_fits_FP_projections} and \ref{Table_results_fits_M-rels}, respectively.

\subsection{Bulge Classification} \label{C/S}
\label{Class_bulges_types}

 We used a classification scheme which segregates pseudobulges from classical bulges according to the following criteria:

\begin{itemize}
\item[I)] Classical bulges fall on the KR (see Fig. 9 of \citetalias{Rios-Lopez_et_al_2021}): pseudobulges are outliers. 

\item[II)] Classical bulges have $n \geq 2$: pseudobulges have low  S\'ersic index with $n < 2$, 

\item [III)] Classical bulges have  $\sigma \geq 130\;\mathrm{km\,s^{-1}}$: pseudobulges have  $\sigma < 130\;\mathrm{km\,s^{-1}}$; this suggests that pseudobulges are less massive than classical bulges. 
\end{itemize}

If at least two of the criteria considered above are met, the bulge is classified either as a classical or a pseudobulge. See \citet{Kormendy-Ho_2013} for a comprehensive review of bulge classification and its limitations.  

\begin{table}
\label{Comp}
\begin{center} 
\caption{Comparison with other works}
\label{Table_comparison_magnitudes}
\begin{tabular}{l l l l}
\hline
Galaxy & $~~M_{K}$ & ~$M_{K_s}$ & ~~~~~~$M_{Ks}$ \\
~~(1) & ~~~(2) & ~~(3) & ~~~~~~~(4)\\
\hline
M~32 & $-19.48$ & $-19.8$ & $-19.45 \pm 0.05$  \\
M~60 &  $-25.90$ & $-25.8$ & $-25.8 \pm 0.05$ \\
M~81  & $---$ & $-24.1$ & $-23.98 \pm 0.03$ \\
M~87  & $-26.08$ & $-25.6$ & $-25.45 \pm 0.05$\\
M~104  & $---$ & $-25.4$ & $-25.13 \pm 0.02$ \\
M~105  & $-24.54$ & $-24.2$ & $-23.92 \pm 0.04$\\
M~106 & $-23.97$ & $-22.4$ & $-23.88 \pm 0.03$  \\
NGC~3115  & $-24.24$ & $-24.4$ & $-24.22 \pm 0.05$ \\
NGC~3377 & $-22.97$ & $-23.6$ & $-22.94 \pm 0.09$  \\
NGC~4697  & $-24.70$ & $-24.6$ & $-24.58 \pm 0.07$ \\
\hline
\multicolumn{4}{l}{(2)  $M_{K}$ from \citet{Lasker_et_al_2014},}\\
\multicolumn{4}{l}{(3) $M_{K_s}$ from \citet{Marconi-Hunt_2003},} \\
\multicolumn{4}{l}{(4) $M_{K_s}$ from this work.}\\
\end{tabular}
\end{center}
\end{table}

\begin{figure}
%\begin{center}
\includegraphics[width=9 cm]{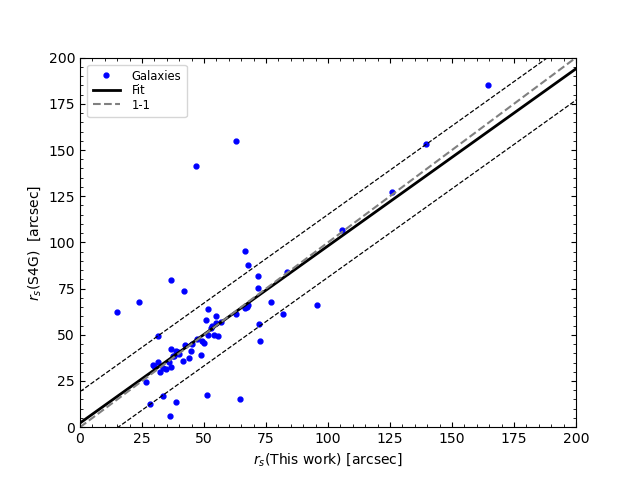}
\caption[rs-plotJ]{\footnotesize Relation between the scale length, $r_s$, in this work using 2MASS data and $r_s$ from the $S^4G$ survey. The solid black line is the linear fit, while the dotted grey line represents a 1-1 line and the dashed lines represent the scatter of the relation.}
\label{rs_plot}
%\end{center}
\end{figure}

\begin{figure*}
\begin{center}
\includegraphics[width=18 cm]{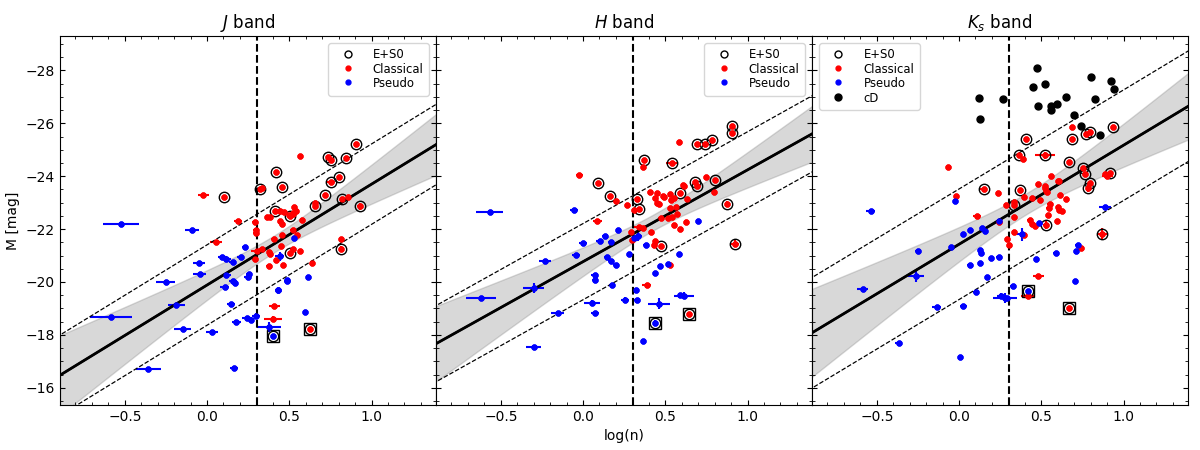}
\caption[n-AbsMag]{\footnotesize Relation between absolute magnitude and S\'ersic index $n$ (LCR) for objects in our sample in the three bands of 2MASS ($J,\, H, \, K_s$). Vertical dashed lines indicate 
$n=2$. Red-filled circles represent classical bulges; the blue ones are pseudobulges according to our classification (\S\ref{Class_bulges_types}). In contrast, the red points inside black circles are the ETGs (indicated by the label E+S0) galaxies. M~32 (classified as classical) and M~110 (classified as pseudo) are highlighted in black squares (see \S5.1 of \citetalias{Rios-Lopez_et_al_2021}). Black dots represent the cD galaxies considered in this work. The black solid line in the $K_s$ band LCR (left panel) fits the whole sample, i.e., including the cD galaxies. The grey-shaded region about the fit corresponds to the $1\sigma$ confidence band, while the dashed black lines indicate the intrinsic scatter of the fit for the entire sample.}
\label{n-AbsMag}
\end{center}
\end{figure*}

\section{Results}
\label{Res}

We compare our results with previous galaxy photometry studies below and present fundamental SRs for structural galaxy parameters, such as the FP and its projections. We have revisited how BH's mass correlates with their host galaxies' physical properties. The distinction between classical and pseudobulges on the SRs is indicated.

\subsection{Comparisons with previous works}\label{sb_phot}

\begin{figure*}
\begin{center}
\includegraphics[width=18 cm]{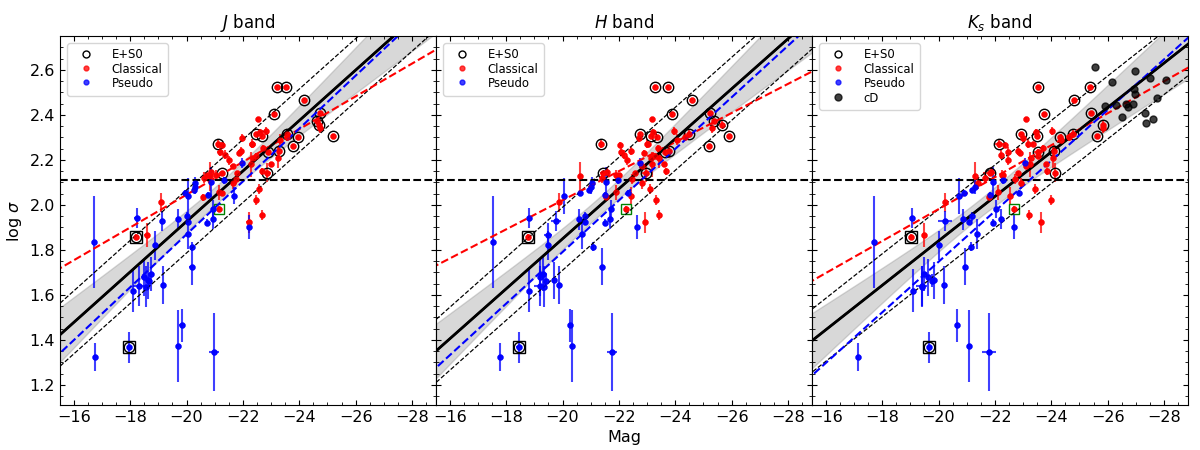}
\caption[FJRfinal]{\footnotesize Faber-Jackson relation (FJR) for galaxies and bulges in our sample in the 2MASS bands. The horizontal black dashed line at $\log \sigma = 2.11$ indicates the division between bulges and pseudobulges. Symbols represent the same as in Fig. \ref{n-AbsMag}, except for the green square indicating the bulge of NGC~4826 (see \S 5.1 of \citetalias{Rios-Lopez_et_al_2021}). The dashed red line represents the fit for classical bulges, while the blue dashed line is for pseudobulges. The black solid line represents the fit of the joint distributions of the whole sample.}
\label{FJR_final}
\end{center}
\end{figure*}

\begin{figure*}
\begin{center}
\includegraphics[width=18 cm]{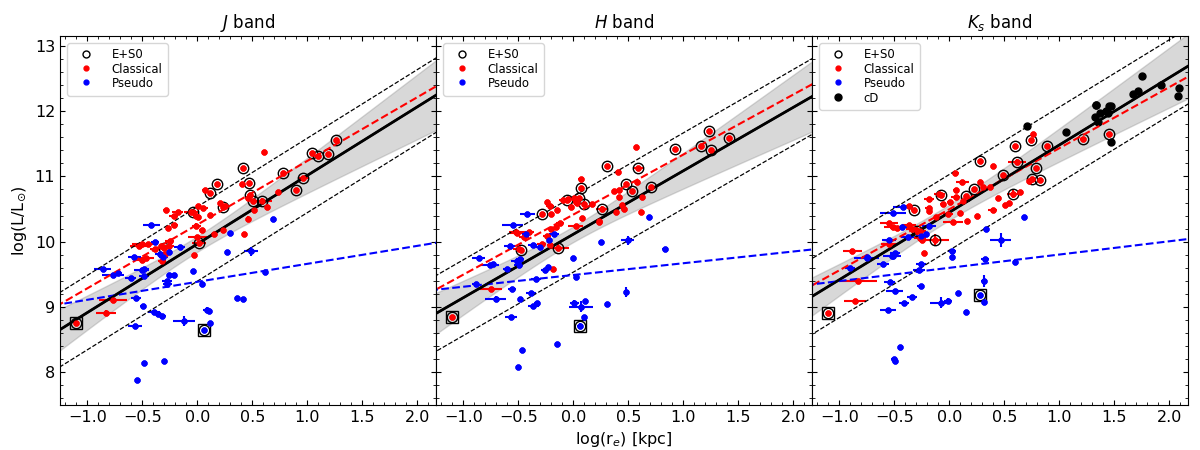}
\caption[LSRfinal]{\footnotesize Luminosity-Size relation (LSR) for galaxies and bulges in our sample for the three bands of 2MASS. Symbols and colours are the same as in previous images. The red dashed line is the linear fit for the classical bulges, the blue dashed line is for pseudobulges, and the black solid line is for the joint distribution of the whole sample.}
\label{LSR_final}
\end{center}
\end{figure*}

We compare measurements for ETGs in our work with those reported by \citet{Marconi-Hunt_2003} and \citet{Lasker_et_al_2014} for the following galaxies in common: M~32 (NGC~221), M~60 (NGC~4649), M~81 (NGC~3031), M~87 (NGC~4486), M~104 (NGC~4594), M~105 (NGC~3379), M~106 (NGC~4258), NGC~3115, NGC~3377 and NGC~4697. Table \ref{Table_comparison_magnitudes} shows the total absolute magnitudes in the $K$ band \citep[from CFHT,][]{Lasker_et_al_2014} and $K_s$ band \citep[from 2MASS,][]{Marconi-Hunt_2003}, shown in column 2 and column 3, respectively, while column 4 shows our total magnitudes. Table \ref{Table_comparison_magnitudes} shows that our measurements are in good agreement with those reported earlier, even though the observations reported by \citet{Lasker_et_al_2014} are four magnitudes deeper than 2MASS.

We want to establish that 2MASS data is right to conduct a detailed SB study of the discs in this study. For this aim, we consider 
$\mu_{K_s}\sim20\; \mathrm{ mag \, arcsec^{-2}}$ corresponds to 
$1\, \sigma$  rms of the background noise approximately \citep{Jarrett_2000,Jarrett_et_al_2000a,Jarrett_2004}, which can be taken as a fiducial SB limit. 
Since we are dealing with the brightest LGA  galaxies ($K_s\leq10\;{\rm mag}$), after modelling bulges, discs and bars (when present), we found that the average central SB for discs is $\mean{\mu_{(0,K_s)}} = 17.8\pm0.1\; \mathrm{mag \, arcsec^{-2}}$ for 71 galaxies considered. This value is consistent with the mean central SB of late-type spirals \citep[Fig. 12, 
Fig. 18 in][respectively]{Jarrett_2000,Jarrett_et_al_2003}. After background subtraction during SB modelling, we reached $\mu_{K_s}\sim 22\; \mathrm{ mag \, arcsec^{-2}}$. Therefore, taking $\mu_{(0,K_s)}=\mean{\mu_{(0,K_s)}}$, we have to stretch to a radius equivalent to four times the scale length, $r=4r_s$, to reach $\mu_{K_s}\sim22\; \mathrm{ mag \, arcsec^{-2}}$; such a radius contains 90\% of the total light of an exponential disc, which is comparable to the total aperture radius $r_t$ introduced by \citet{Jarrett_et_al_2003}. Secondly, we consider the work of \citet{2015ApJS..219....4S}, who used \texttt{GALFIT} to model galaxies from the Spitzer $S^4G$ survey \citep{2010PASP..122.1397S}. We chose to compare the models in the Spitzer $3.6\, \mu m$ band for the 63 ETG in common with \citet{2015ApJS..219....4S}, which closely samples the same stellar populations as the $K_s$ band. Despite the variations in depth, data processing and reduction, and fitting strategies, the fitted scale lengths from each study agree, as the average $r_s$ ratio is close to unity:  
\begin{equation} \label{dif}
 \mean{\frac{r_s(3.6\,\mu m)}{r_s(K_s)}} =1.11\pm0.06.
\end{equation}

Similarly, from a direct comparison between the scale length $r_s$ measured in this work and the one from the $S^4G$ survey, we found a slope of $m_{(\rm S^4G\,vs.\,2MASS)}=0.96 \pm 0.08$ from the linear regression analysis (see Fig. \ref{rs_plot}), which is consistent with the mean ratio reported in Eq. \ref{dif}. We consider that a more detailed SB analysis lies beyond the scope of this work; hence,  we leave it for a forthcoming publication. Therefore, we can safely conclude that 2MASS data reach sufficient depth and S/N to model the SB of the brightest nearby spiral galaxies accurately.

\citet{2004ApJ...617..879L} studied a large sample of cD galaxies using corrected 2MASS SXC  \citep{Jarrett_et_al_2000a} magnitudes. However, our $K_s$ band total magnitudes are $\sim0.5\, \mathrm {mag}$ brighter for the galaxies in common. Our result agrees with the studies of \citet{Lauer_et_al_2007} and \citet{2024MNRAS.527..249Q}.

\subsection{Galaxy Scaling Relations}\label{Scal_rels_Res}

\begin{table*}
\caption{\textbf{Galaxy Scaling Relations.} A correlation of the form $y = m(x - \langle x \rangle) + b + \epsilon_y$ is assumed. Columns (1), (2) $\&$ (3) are for the scaling relation, (sub)sample and data points, respectively, used in the fit: \textit{All} for the whole sample, \textit{C} for classical bulges, \textit{P} for pseudobulges, \textit{All+cD} for all the sample plus the cD galaxies and \textit{cD} for the subsample of cD galaxies. (4) $\&$ (5) Slope ($m$) and zero point ($b$) of the regression line fit, (6) intercept ($b_o$). (7) Intrinsic scatter of the relation. (8) Mean value of the abscissae ${x_i}$. (9) Pearson correlation coefficient.}
\label{Table_results_fits_FP_projections}
\begin{center}
\tabcolsep 5pt
\begin{tabular}{c c c c c c c c c }
\hline
Relation & Group & $N$ & \textbf{$m$} & \textbf{$b$} & \textbf{$b_o$} & \textbf{$\epsilon_y$} & $\langle x \rangle$ &\textbf{$r$} \\
\& Band \\
\vspace{0.01cm}\\
(1) & (2) & (3) & (4) & (5) & (6) & (7) & (8) & (9) \\
\hline
\multicolumn{9}{c}{Luminosity-Concentration Relation (LCR)}\\
LCR $|$ $J$  & All    & 101 & $-3.81 \pm 0.57$ & $-21.45 \pm 0.16$ & $-19.89\pm 0.28$ & 1.52 $\pm$ 0.13 &   0.41   & $-0.50$ \\
             & C      & 60  & $-2.06 \pm 0.70$ & $-22.38 \pm 0.15$ & $-21.43\pm 0.36$ & 1.09 $\pm$ 0.12 &   0.46   & $-0.30$ \\
             & P      & 41  & $-0.38 \pm 0.70$ & $-19.88 \pm 0.19$ & $-19.82 \pm 0.22$ & 1.15 $\pm$ 0.17 &   0.17   & $-0.14$ \\
LCR $|$ $H$  & All    & 101 & $-3.47 \pm 0.53$ & $-22.28 \pm 0.15$ & $-20.79 \pm 0.29 $ & 1.44 $\pm$ 0.12 &   0.43   & $-0.46$ \\
             & C      & 59  & $-2.12 \pm 0.60$ & $-22.92 \pm 0.12$ & $-21.80\pm 0.34 $ & 0.82 $\pm$ 0.10 &   0.53   & $-0.26$ \\
             & P      & 42  & $-0.05 \pm 0.70$ & $-20.40 \pm 0.21$ & $-20.39\pm 0.24 $ & 1.31 $\pm$ 0.19 &   0.18   & $-0.11$ \\
LCR $|$ $K_s$& All    & 101 & $-2.86 \pm 0.49$ & $-22.46 \pm 0.15$ & $-21.26\pm 0.27$ & 1.49 $\pm$ 0.12 &   0.42   & $-0.50$ \\
             & C      & 61  & $-2.72 \pm 0.63$ & $-23.47 \pm 0.13$ & $-22.03\pm 0.37$ & 0.96 $\pm$ 0.11 &   0.53   & $-0.25$ \\
             & P      & 40  & $-0.29 \pm 0.60$ & $-20.90 \pm 0.20$ & $-20.86\pm 0.22$ & 1.13 $\pm$ 0.18 &   0.15   & $-0.21$ \\
             & All+cD & 119 & $-3.74 \pm 0.62$ & $-23.18 \pm 0.20$ & $-21.42\pm 0.37$ & 2.09 $\pm$ 0.16 &   0.47   & $-0.49$ \\
             & cD     & 18  & $-1.90 \pm 0.43$ & $-26.76 \pm 0.09$ & $-25.66\pm 0.29$ & 0.43 $\pm$ 0.32 &   0.58   & $-0.38$ \\
\hline 
 \multicolumn{9}{c}{Faber-Jackson Relation (FJR)}\\

FJR $|$ $J$  & All    & 97  & $-0.11 \pm 0.01$ & 2.07 $\pm$ 0.02   & $-0.27\pm 0.21$ & 0.14 $\pm$ 0.01 & $-21.29$ & $-0.80$ \\
             & C      & 60  & $-0.07 \pm 0.01$ & 2.21 $\pm$ 0.01   & $0.65 \pm 0.22$ & 0.10 $\pm$ 0.01 & $-22.27$ & $-0.71$ \\
             & P      & 37  & $-0.12 \pm 0.02$ & 1.83 $\pm$ 0.03   & $-0.54\pm0.40$ & 0.17 $\pm$ 0.03 & $-19.71$ & $-0.56$ \\
FJR $|$ $H$  & All    & 97  & $-0.11 \pm 0.01$ & 2.07 $\pm$ 0.02   & $-0.35$ $\pm$ 0.22 & 0.14 $\pm$ 0.01 & $-21.99$ & $-0.78$ \\
             & C      & 59  & $-0.06 \pm 0.01$ & 2.22 $\pm$ 0.02   & $~~0.84 \pm 0.23$ & 0.11 $\pm$ 0.01 & $-23.01$ & $-0.61$ \\
             & P      & 38  & $-0.12 \pm 0.02$ & 1.84 $\pm$ 0.03   & $-0.61$ $\pm$ 0.41 & 0.17 $\pm$ 0.03 & $-20.4$  & $-0.54$ \\
FJR $|$ $K_s$& All    & 97  & $-0.11 \pm 0.01$ & 1.85 $\pm$ 0.02   & $-0.39$ $\pm$ 0.22 & 0.14 $\pm$ 0.01 & $-20.35$ & $-0.74$ \\
             & C      & 61  & $-0.07 \pm 0.01$ & 1.99 $\pm$ 0.04   & $0.58\pm 0.23$ & 0.10 $\pm$ 0.01 & $-20.14$ & $-0.67$ \\
             & P      & 36  & $-0.11 \pm 0.03$ & 1.83 $\pm$ 0.03   & $-0.45 \pm 0.62$ & 0.17 $\pm$ 0.03 & $-20.72$ & $-0.55$ \\
             & All+cD & 114 & $-0.10 \pm 0.01$ & 1.97 $\pm$ 0.02   & $-0.16$ $\pm$ 0.23 & 0.14 $\pm$ 0.01 & $-21.33$ & $-0.83$ \\
             & cD     & 17  & $ 0.02 \pm 0.01$ & 2.46 $\pm$ 0.03   & $~~2.98$ $\pm$ 0.26 & 0.12 $\pm$ 0.04 & $-25.77$ & 0.45    \\
 \hline
 \multicolumn{9}{c}{Luminosity-Size Relation (LSR)}\\

LSR $|$ $J$  & All    & 101 & 1.05 $\pm$ 0.12  & 9.95 $\pm$ 0.06   & $9.97 $ $\pm$ 0.08 & 0.57 $\pm$ 0.05 & $-0.02$  & 0.67    \\
             & C      & 60  & 0.98 $\pm$ 0.07  & 9.99 $\pm$ 0.04   & $9.87 $ $\pm$ 0.08 & 0.27 $\pm$ 0.03 &   0.12   & 0.88    \\
             & P      & 41  & 0.28 $\pm$ 0.23  & 8.94 $\pm$ 0.09   & $9.00 $ $\pm$ 0.10 & 0.56 $\pm$ 0.08 & $-0.21$  & 0.19    \\
LSR $|$ $H$  & All    & 101 & 0.97 $\pm$ 0.12  & 10.09$\pm$ 0.06   & $10.12$ $\pm$ 0.08 & 0.58 $\pm$ 0.05 & $-0.03$  & 0.64    \\
             & C      & 59  & 0.92 $\pm$ 0.07  & 9.97 $\pm$ 0.04   & $9.85 $ $\pm$ 0.07 & 0.27 $\pm$ 0.03 &   0.13   & 0.86    \\
             & P      & 42  & 0.11 $\pm$ 0.21  & 8.90 $\pm$ 0.09   & $8.93 $ $\pm$ 0.10 & 0.56 $\pm$ 0.08 & $-0.25$  & 0.09    \\
LSR $|$ $K_s$& All    & 101 & 0.96 $\pm$ 0.12  & 10.18$\pm$ 0.06   & 10.21 $\pm$ 0.08 & 0.58 $\pm$ 0.05 & $-0.03$  & 0.63    \\
             & C      & 61  & 0.92 $\pm$ 0.07  & 9.74 $\pm$ 0.04   & 9.65 $\pm$ 0.07 & 0.27 $\pm$ 0.03 &   0.10   & 0.86    \\
             & P      & 40  & 0.15 $\pm$ 0.24  & 8.72 $\pm$ 0.09   & 8.75 $\pm$ 0.10 & 0.58 $\pm$ 0.08 & $-0.22$  & 0.10    \\
             & All+cD & 119 & 0.96 $\pm$ 0.11  & 10.40$\pm$ 0.06   & 10.21 $\pm$ 0.13 & 0.54 $\pm$ 0.04 &   0.20   & 0.82    \\
             & cD     & 18  & 0.77 $\pm$ 0.21  & 10.77$\pm$ 0.11   & 9.68 $\pm$ 0.32 & 0.44 $\pm$ 0.12 &   1.42   & 0.73    \\
 \hline
 \multicolumn{9}{c}{Colour-Magnitude Relation (CMR) for the discs}\\

CMR $|$ $(J-H)\,vs.\, H$     & All    & 85  & $-0.002\pm0.011$ & $0.638 \pm 0.013$ & $~~0.64 $ $\pm$ 0.25  & $0.100\pm0.011$ & $-23.02$ & $-0.25$ \\
~~CMR $|$ $(H-K_s)\,vs.\, H$ & All    & 85  & $-0.006\pm0.015$ & $0.190 \pm 0.016$ & $-0.04$ $\pm$ 0.34 & $0.137\pm0.014$ & $-23.01$ &   0.092 \\
~CMR $|$ $(J-K_s)\,vs.\, H$  & All    & 85  & $-0.022\pm0.014$ & $0.824 \pm 0.016$ & $~~0.36 $ $\pm$ 0.32 & $0.137\pm0.014$ & $-23.02$ & $-0.19$ \\
\hline
\multicolumn{9}{c}{Tully-Fisher Relation (TFR)}\\

TFR $|$ $J$                  & All    & 85  & 0.25 $\pm$ 0.03  & 2.25 $\pm$ 0.01   & $-0.40$ $\pm$ 0.32  & 0.12 $\pm$ 0.01 & 10.60    & 0.64    \\
~TFR $|$ $H$                 & All    & 85  & 0.26 $\pm$ 0.03  & 2.26 $\pm$ 0.01   & $-0.53$ $\pm$ 0.32 & 0.11 $\pm$ 0.01 & 10.74    & 0.62    \\
~~TFR $|$ $K_s$              & All    & 85  & 0.26 $\pm$ 0.03  & 2.24 $\pm$ 0.01   & $-0.45$ $\pm$ 0.32 & 0.12 $\pm$ 0.01 & 10.78    & 0.64    \\
\hline
\multicolumn{9}{c}{Luminosity-Scale Length Relation (LSR$_d$)}\\

LSR$_d$ $|$ $J$              & All    & 85  & $1.32 \pm 0.13$ & $10.06 \pm 0.03 $ & $ 9.43 \pm 0.08$ & $0.27 \pm 0.02$ & 0.48 & 0.63  \\
~LSR$_d$ $|$ $H$             & All    & 85  & $1.35 \pm 0.13$ & $10.02 \pm 0.03 $ & $ 9.38 \pm 0.08$ & $0.27 \pm 0.02$ & 0.47 & 0.64  \\
~~LSR$_d$ $|$ $K_s$          & All    & 85  & $1.39 \pm 0.13$ & $9.78  \pm 0.03 $ & $ 9.13 \pm 0.08$ & $0.26 \pm 0.02$ & 0.47 & 0.72  \\
\hline
\end{tabular}
\end{center}
\end{table*}

\begin{table*}
\caption{\textbf{SMBH Scaling Relations.} A correlation of the form $\logd(\frac{M_{\bullet}}{M_{\sun}}) = m_{\bullet} (\logd x - \langle \logd x \rangle) + b_{\bullet} + \epsilon_{\bullet}$ is assumed. Columns: (1), (2) \& (3) Variable, (sub)sample and data points, respectively, used in the fit: \textit{All} for the whole sample, \textit{C} for classical bulges, \textit{P} for pseudobulges, and * indicates that pseudobulges (not in our sample) from literature were added. (4) \& (5) are the slope and zero point, while (6) is the intercept ($b_{\circ_\bullet})$ of the regression line. (7) Intrinsic scatter of the relation. (8) Mean value of abscissae ($x_i$). (9) Pearson correlation coefficient.}
\label{Table_results_fits_M-rels}
\begin{center}
\tabcolsep 5pt
\begin{tabular}{c c c c c c c c c}
\hline
Variable & Group & $N$ & \textbf{$m_{\bullet}$} & \textbf{$b_{\bullet}$} & \textbf{$b_{\circ_\bullet}$} & \textbf{$\epsilon_{\bullet}$} & $\langle \logd x \rangle$ & \textbf{$r$} \\
(1) & (2) & (3) & (4) & (5) & (6) & (7) & (8) & (9)\\
 \hline
 \hline
 \multicolumn{9}{c}{Bulge Luminosity}\\

$L_{J}$ & All & 33 & 1.40 $\pm$ 0.21 & 7.85 $\pm$ 0.12 & $-5.84 \pm 2.06$& 0.64 $\pm$ 0.11 & 9.78 & 0.78 \\
 & C & 25 & 1.38 $\pm$ 0.29 & 8.11 $\pm$ 0.16 & $-5.63 \pm 2.90$ & 0.73 $\pm$ 0.16 & 9.96 & 0.71 \\
 & P & 8 & 2.28 $\pm$ 0.49 & 7.38 $\pm$ 0.10 & $-13.71 \pm 4.54$ & 0.12 $\pm$ 0.10 & 9.25 & 0.78 \\
$L_{H}$ & All & 33 & 1.29 $\pm$ 0.23 & 7.85 $\pm$ 0.13 & $-4.69 \pm 2.24 $& 0.71 $\pm$ 0.12 & 9.72 & 0.73 \\
 & C & 25 & 1.23 $\pm$ 0.32 & 8.11 $\pm$ 0.17 & $-4.07 \pm 3.18 $ & 0.81 $\pm$ 0.17 & 9.90 & 0.64 \\
 & P & 8 & 1.83 $\pm$ 0.38 & 7.49 $\pm$ 0.10 & $-9.35 \pm 3.50 $  & 0.11 $\pm$ 0.09 & 9.20 & 0.75 \\
$L_{Ks}$ & All & 35 & 1.38 $\pm$ 0.18 & 8.00 $\pm$ 0.12 & $-6.48 \pm 1.90 $ & 0.68 $\pm$ 0.11 & 10.49 & 0.80 \\
 & C & 27 & 1.33 $\pm$ 0.24 & 8.29 $\pm$ 0.16 & $-5.91 \pm 2.57 $ & 0.77 $\pm$ 0.15 & 10.68 & 0.77 \\
 & P & 8 & 2.08 $\pm$ 0.44 & 7.44 $\pm$ 0.09 & $-11.40 \pm 3.99 $ & 0.24 $\pm$ 0.17 & 9.06 & 0.76 \\
$L_{K_s*}$ & All & 43 & 1.18 $\pm$ 0.23 & 7.70 $\pm$ 0.12 & $-3.47 \pm 2.18$ & 0.75 $\pm$ 0.11 & 10.25 & 0.65 \\
 & P & 16 & 0.30 $\pm$ 0.31 & 7.12 $\pm$ 0.12 & $4.37 \pm 2.84$ & 0.46 $\pm$ 0.15 & 9.16 & 0.33 \\
 \hline
 \multicolumn{9}{c}{Total Galaxy Luminosity}\\

$L_{tot,J}$ & All & 33 & 1.27 $\pm$ 0.29 & 7.86 $\pm$ 0.15 & $-5.32 \pm 3.02$ & 0.81 $\pm$ 0.14 & 10.38 & 0.63 \\
 & C & 25 & 1.12 $\pm$ 0.36 & 8.11 $\pm$ 0.19 & $ -3.63 \pm 3.78 $ & 0.88 $\pm$ 0.19 & 10.48 & 0.55 \\
 & P & 8 & 4.60 $\pm$ 1.40 & 5.74 $\pm$ 0.66 & $ -40.77 \pm 14.18 $ & 0.13 $\pm$ 0.12 & 10.11 & 0.73 \\
$L_{tot,H}$ & All & 33 & 1.29 $\pm$ 0.29 & 7.86 $\pm$ 0.15 & $-5.45 \pm 3.00$  & 0.80 $\pm$ 0.14 & 10.32 & 0.64 \\
 & C & 25 & 1.15 $\pm$ 0.36 & 8.11 $\pm$ 0.18 & $-3.87 \pm 3.76 $ & 0.87 $\pm$ 0.18 & 10.42 & 0.57 \\
 & P & 8 & 4.40 $\pm$ 1.10 & 6.23 $\pm$ 0.24 & $-37.95 \pm 11.06 $ & 0.14 $\pm$ 0.12 & 10.04 & 0.74 \\
$L_{tot,K_s}$ & All & 35 & 1.44 $\pm$ 0.26 & 8.01 $\pm$ 0.15 & $-6.66 \pm 2.66 $& 0.83 $\pm$ 0.14 & 10.19 & 0.71 \\
 & C & 27 & 1.34 $\pm$ 0.32 & 8.29 $\pm$ 0.18 & $-5.51 \pm 3.30 $ & 0.89 $\pm$ 0.18 & 10.30 & 0.66 \\
 & P & 8 & 4.12 $\pm$ 0.98 & 6.32 $\pm$ 0.19 & $-34.26 \pm 9.68 $ & 0.30 $\pm$ 0.13 & 9.85 & 0.72 \\
$L_{tot,K_s*}$ & All & 53 & 0.10 $\pm$ 0.03 & 7.56 $\pm$ 0.13 & $6.74 \pm 0.28$ & 0.87 $\pm$ 0.11 & 8.24 & 0.38 \\
 & P & 26 & 0.023 $\pm$ 0.02 & 7.05 $\pm$ 0.09 & $6.90 \pm 0.16 $ & 1.42 $\pm$ 0.10 & 6.50 & 0.20 \\
 \hline
 \multicolumn{9}{c}{Effective Radius ($r_e$)}\\

$r_{e,J}$ & All & 31 & 1.50 $\pm$ 0.37 & 7.86 $\pm$ 0.15 & $ 8.01 \pm 0.19$ & 0.84 $\pm$ 0.14 & -0.10 & 0.60 \\
 & C & 23 & 1.34 $\pm$ 0.49 & 8.11 $\pm$ 0.19 &  8.07 $\pm$ 0.22 & 0.91 $\pm$ 0.19 & 0.03 & 0.51 \\
  & P & 8 & 1.92 $\pm$ 0.98 & 7.54 $\pm$ 0.11 &  8.46 $\pm$ 0.49 & 0.64 $\pm$ 0.38 & -0.48 & 0.28 \\
$r_{e,H}$ & All & 31 & 1.23 $\pm$ 0.37 & 7.86 $\pm$ 0.16 & 8.02 $\pm$ 0.19 & 0.88 $\pm$ 0.15 & -0.13 & 0.53 \\
 & C & 23 & 1.02 $\pm$ 0.51 & 8.11 $\pm$ 0.21 & 8.08 $\pm$ 0.23 & 0.97 $\pm$ 0.20 & 0.03 & 0.40 \\
 & P & 8 & 0.70 $\pm$ 1.40 & 7.12 $\pm$ 0.23 & 7.52 $\pm$ 0.83 & 0.60 $\pm$ 0.38 & -0.57 & 0.26 \\
$r_{e,Ks}$ & All & 31 & 1.17 $\pm$ 0.39 & 7.86 $\pm$ 0.17 & $8.04 \pm 0.20$ & 0.89 $\pm$ 0.16 & -0.15 & 0.48 \\
 & C & 23 & 0.95 $\pm$ 0.52 & 8.11 $\pm$ 0.21 & $ 8.12 \pm 0.22 $ & 0.98 $\pm$ 0.21 & -0.01 & 0.37 \\
 &P & 8 & 1.26 $\pm$ 0.98 & 7.22 $\pm$ 0.27 & $ 7.81 \pm 0.54 $ & 0.71 $\pm$ 0.44 & -0.47 & 0.22 \\
$r_{e,K_s*}$ & All & 41 & 1.25 $\pm$ 0.30 & 7.64 $\pm$ 0.13 & $7.95 \pm 0.17$ & 0.83 $\pm$ 0.12 & -0.25 & 0.55 \\
 & P & 18 & 0.37 $\pm$ 0.41 & 7.02 $\pm$ 0.11 & $ 7.22 \pm 0.25 $ & 0.44 $\pm$ 0.13 & -0.55 & 0.25 \\
 \hline
 \multicolumn{9}{c}{Sérsic Index($n$)}\\

$n_{J}$ & All & 31 & -0.09 $\pm$ 0.89 & 7.86 $\pm$ 0.19 & $7.90 \pm 0.43 $ & 1.05 $\pm$ 0.18 & 0.43 & $-0.17$ \\
 & C & 23 & -0.90 $\pm$ 1.10 & 8.12 $\pm$ 0.22 & $ 8.54 \pm 0.56 $ & 1.05 $\pm$ 0.22 & 0.47 & $-0.17$ \\
 & P & 8 & -1.00 $\pm$ 1.20 & 7.11 $\pm$ 0.22 & $ 7.41 \pm 0.43 $ & 0.57 $\pm$ 0.35 & 0.30 & $-0.40$ \\
$n_{H}$ & All & 31 & -0.33 $\pm$ 0.79 & 7.86 $\pm$ 0.19 & $ 8.00 \pm 0.40 $ & 1.05 $\pm$ 0.18 & 0.43 & $-0.08$ \\
 & C & 23 & -1.46 $\pm$ 0.94 & 8.12 $\pm$ 0.21 & $ 8.82 \pm 0.50 $ & 1.01 $\pm$ 0.21 & 0.48 & $-0.31$ \\
 & P & 8 & -2.7 $\pm$ 0.86 & 6.76 $\pm$ 0.10 & $ 7.38 \pm 0.0.28 $ & 0.30 $\pm$ 0.16 & 0.23 & $-0.18$ \\
$n_{K_s}$ & All & 31 & -0.37 $\pm$ 0.70 & 7.86 $\pm$ 0.19 & $ 8.02 \pm 0.35 $ & 1.05 $\pm$ 0.18 & 0.42 & $-0.10$ \\
 & C & 23 & -1.11 $\pm$ 0.85 & 8.12 $\pm$ 0.22 & $ 8.64 \pm 0.46 $ & 1.02 $\pm$ 0.21 & 0.47 & $-0.27$ \\
 & P & 8 & -2.36 $\pm$ 0.82 & 6.78 $\pm$ 0.17 & $ 7.49 \pm 0.39 $ & 0.30 $\pm$ 0.27 & 0.30 & $-0.19$ \\
 \hline
 \multicolumn{9}{c}{Central Velocity Dispersion ($\sigma$)}\\

$\sigma$ & All & 33 & 5.45 $\pm$ 0.52 & 8.01 $\pm$ 0.10 & $-4.03 \pm 1.16 $ & 0.51 $\pm$ 0.10 & 2.21 & 0.88 \\
 & C & 25 & 5.74 $\pm$ 0.69 & 8.31 $\pm$ 0.12 & $-4.72 \pm 1.58 $ & 0.56 $\pm$ 0.13 & 2.27 & 0.86 \\
 & P & 8 & 5.50 $\pm$ 1.80 & 7.08 $\pm$ 0.17 & $-4.03 \pm 3.64 $ & 0.11 $\pm$ 0.10 & 2.02 & 0.81 \\
 $\sigma_*$ & All & 47 & 5.46 $\pm$ 0.54 & 7.74 $\pm$ 0.09 & $-4.22 \pm 1.19 $ & 0.58 $\pm$ 0.09 & 2.19 & 0.82 \\
 & P & 22 & 1.69 $\pm$ 0.92 & 7.06 $\pm$ 0.10 & $3.51 \pm 1.94$ & 0.41 $\pm$ 0.11 & 2.10 & 0.42 \\
 \hline
\end{tabular}
\end{center}
\end{table*}

% CONTINUED TABLE 5
\begin{table*}
\contcaption{}
\begin{center}
\tabcolsep 5pt
\begin{tabular}{c c c c c c c c c}
\hline
Variable & Group & $N$ & \textbf{$m_{\bullet}$} & \textbf{$b_{\bullet}$} & \textbf{$b_{\circ_\bullet}$} & \textbf{$\epsilon_{\bullet}$} & $\langle \logd x \rangle$ & \textbf{$r$} \\

(1) & (2) & (3) & (4) & (5) & (6) & (7) & (8) & (9) \\
\hline
\multicolumn{9}{c}{Velocity Dispersion at the effective radius ($\sigma_{e}$)}\\

$\sigma_{e}$ & All & 24 & 5.72 $\pm$ 0.49 & 8.00 $\pm$ 0.08 & $-4.76 \pm 1.11 $ & 0.34 $\pm$ 0.10 & 2.23 & 0.91 \\
 & C & 20 & 5.69 $\pm$ 0.55 & 8.14 $\pm$ 0.09 & $ -4.72 \pm 1.26 $ & 0.37 $\pm$ 0.12 & 2.26 & 0.90 \\
& P & 4 & 2.90 $\pm$ 3.50 & 7.52 $\pm$ 0.18 & $ 1.37 \pm 7.42 $ & 0.23 $\pm$ 0.10 & 2.12 & 0.59\\
$\sigma_{e*}$ & All & 42 & 5.26 $\pm$ 0.45 & 7.62 $\pm$ 0.07 & $ -3.79 \pm 0.99 $ & 0.41 $\pm$ 0.08 & 2.17 & 0.85 \\
 & P & 22 & 1.83 $\pm$ 0.94 & 7.09 $\pm$ 0.09 & $ 3.25 \pm 1.98 $ & 0.33 $\pm$ 0.10 & 2.10 & 0.36\\
\hline
 \multicolumn{9}{c}{Rotation Velocity ($V_{rot}$)}\\
$V_{rot}$& All & 24 & 2.70 $\pm$ 1.4 & 7.68 $\pm$ 0.19 &  $1.44 \pm 3.24$  & 0.90 $\pm$ 0.18 & 2.31 & 0.38 \\
 & C & 16 & 5.30 $\pm$ 1.7 & 7.45 $\pm$ 0.13 & $-4.85 \pm 4.02$ & 0.30 $\pm$ 0.20 & 2.32 & 0.24 \\
 & P & 8 & 4.10 $\pm$ 1.60 & 7.19 $\pm$ 0.21 & $-2.24 \pm 3.69$ & 0.54 $\pm$ 0.33 & 2.30 & 0.74\\
$V_{rot*}$ & All & 30 & 2.40 $\pm$ 1.30 & 7.58 $\pm$ 0.16 & $ 2.01 \pm 3.02$ & 0.86 $\pm$ 0.15 & 2.32 & 0.33\\
 & P & 14  & 3.70 $\pm$ 1.40 & 7.20 $\pm$ 0.15 & $ -1.38 \pm 3.25$ & 0.47 $\pm$ 0.18 & 2.32 & 0.60\\
\hline
\end{tabular}
\end{center}
\end{table*}

\subsubsection{Luminosity-Concentration Relation (LCR)}

The Luminosity-Concentration Relation \citep[LCR; e.g.,][]{Graham_2001, Fisher-Drory_2008} is shown in Fig. \ref{n-AbsMag}, involving absolute magnitude for galaxies and bulges and the S\'ersic index $n$ of the bulge. The dashed vertical line in Fig. \ref{n-AbsMag} indicates $n=2$, one of the criteria for discriminating between bulge types used in this work. According to our classification, red circles are for classical bulges, while the blue ones are for pseudobulges (\S\ref{Class_bulges_types}). In contrast, the ETG (indicated by the label E+S0) are red points inside black circles, and the black dots are cD galaxies. More luminous galaxies tend to have higher concentration indexes.

\subsubsection{Faber-Jackson Relation (FJR)} \label{FJR}

The FJR shows an interplay between the central stellar $\sigma$ and galaxies' absolute magnitude or luminosity. The FJR is depicted in Fig. \ref{FJR_final}, where the dashed horizontal line at  $\log \sigma = 2.11$ (corresponding to $\sigma = 130 \; \mathrm{km \,s^{-1}}$) indicates our criterion adopted to separate bulge types. The black line is the fit for the whole sample, while the red and blue dotted lines are for classical and pseudo bulges, respectively. M~32 (classified as classical) and M~110 (classified as pseudo) are highlighted in black squares (see \S5.1 of \citetalias{Rios-Lopez_et_al_2021}). We also see that classical bulges and pseudobulges have different trends as indicated by their linear fits; however, by considering E+S0, cD galaxies, classical bulges and pseudobulges, a correlation can be formed with a large scatter dominated by pseudobulges. 

\subsubsection{Luminosity-Size Relation (LSR)}

The LSR generated for our sample is shown in Fig. \ref{LSR_final} for the 2MASS bands, where the linear fits and the colour code are the same as previously. Our results show that galaxies, regardless of the environment, follow the LSR; pseudobulges display more significant dispersion and seem only weakly correlated, showing flat slopes.

\subsubsection{The Fundamental Plane (FP)}

The FP correlates the size with galaxies' average SB and $\sigma$ of galaxies. As we said before, the FP and its projections have been used as distance indicators, as well as in the study of galaxy formation and evolution \citep[e.g.,][]{Dressler_et_al_1987, Djorgovski-Davis_1987, Cappellari_2015, Saulder_et_al_2019}.

\begin{figure*}
\begin{center}
\includegraphics[width=18 cm]{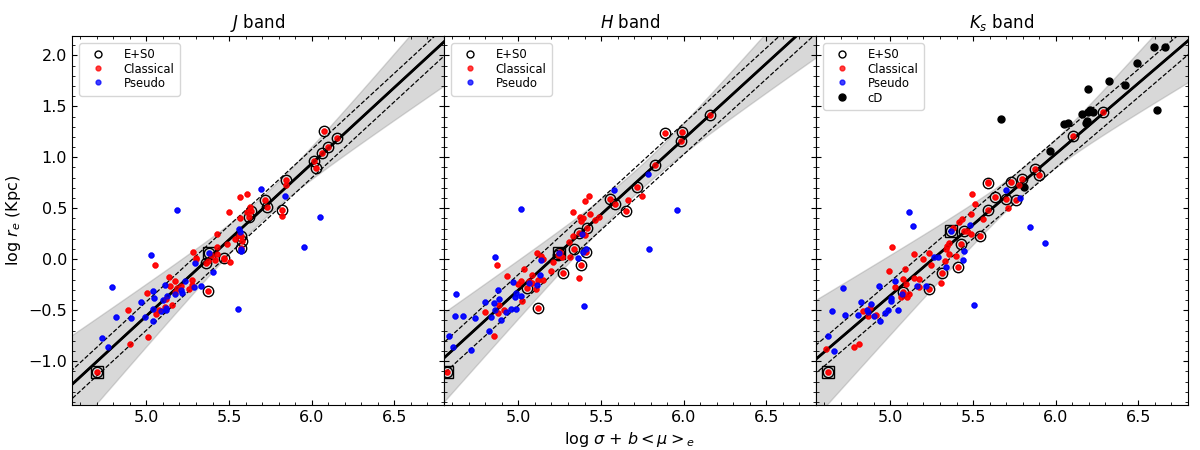}
\caption[FP-final]{\footnotesize The Fundamental Plane (FP) for galaxies and bulges of our sample. The three bands of 2MASS are shown. The black line is the linear fit for the whole sample. Symbols and colours represent the same as in previous images.}
\label{FP_final}
\end{center}
\end{figure*}

Here, we constructed our FP for E+S0, bulges, and cD galaxies (only in $K_s$ band) shown in Fig. \ref{FP_final}, where symbols are coded the same way as previously. We have obtained the following fits for the FP for each 2MASS band, given by Equations \ref{FP1}, \ref{FP2} and \ref{FP3}, below:
{\small
\begin{align}
\nonumber\logd r_{e}(J) &= (1.49 \pm 0.06) \logd \sigma + (0.28 \pm 0.01)\mean{\mu_{e}(J)}-\\ \label{FP1}
&-(8.03 \pm 0.19),\\ 
\nonumber \logd r_{e}(H) &= (1.48 \pm 0.06) \logd \sigma + (0.28 \pm 0.01) \mean{\mu_{e}(H)}-\\ \label{FP2}
&-(7.71 \pm 0.18),\\ 
\nonumber\logd r_{e}(K_s) &= (1.39 \pm 0.07) \logd \sigma + (0.27 \pm 0.01)\mean{\mu_{e}(K_s)}-\\
&-(7.29 \pm 0.21). \label{FP3}
\end{align}
}

\subsubsection{Colour-Magnitude Relation (CMR)}\label{Col-Mag}
\begin{figure}
\begin{center}
\includegraphics[width=8.35 cm]{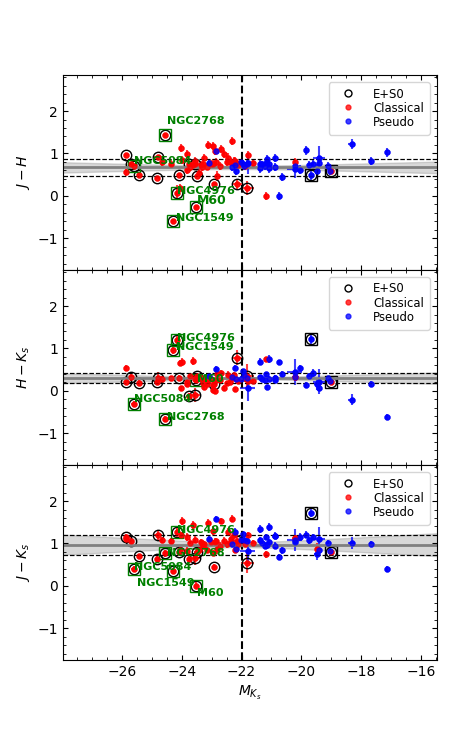}
\caption[CMR]{\footnotesize Colour-Magnitude Relation (CMR) for colours $J-H$ (top panel), $H-K_s$ (middle panel), and $J-K_s$ (bottom panel), while on the x-axes we plot the total bulge luminosity in the $Ks$ 
band. Symbols and colours are the same as in previous images. The grey solid line represents the mean colour, while the dashed vertical line represents a new bulge/pseudobulge classifier: classical bulges are more luminous ($M_{K_s}\leq -22$) than pseudobulges. Green squares indicate sources with AGN or peculiar classification (see \S\ref{Dis_gal_scal_rel}).}
\label{CMR}
\end{center}
\end{figure}

\begin{figure}
\begin{center}
\includegraphics[width=8.5 cm]{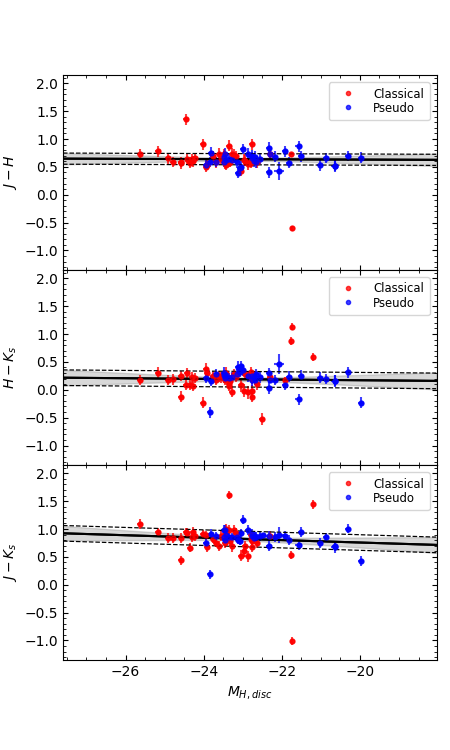}
\caption[CMR-discs]{\footnotesize CMR for galaxies discs: disc colours vs. total disc luminosity. Symbols and colours are the same as in previous images, but on the abscissa, we have used the absolute magnitude of the discs in the $H$ band: $M_{H, disc}$.}
\label{CMR_discs}
\end{center}
\end{figure}

The CMR diagrams for bulges and discs are presented in Figs. \ref{CMR} and \ref{CMR_discs} for the colours $J-H$, $H-K_s$ and $J-K_s$. We plot the total magnitude of the bulge and disc components on the x-axis, respectively. Colours were generated using the component's total magnitude in each band. We found that the colours of bulges and pseudobulges are indistinguishable, but classical bulges are more luminous than pseudobulges. Nevertheless, unlike bulges, we found that more luminous discs are slightly redder, conforming to a CMR in the NIR.

\begin{figure*}
\begin{center}
\includegraphics[width=18 cm]{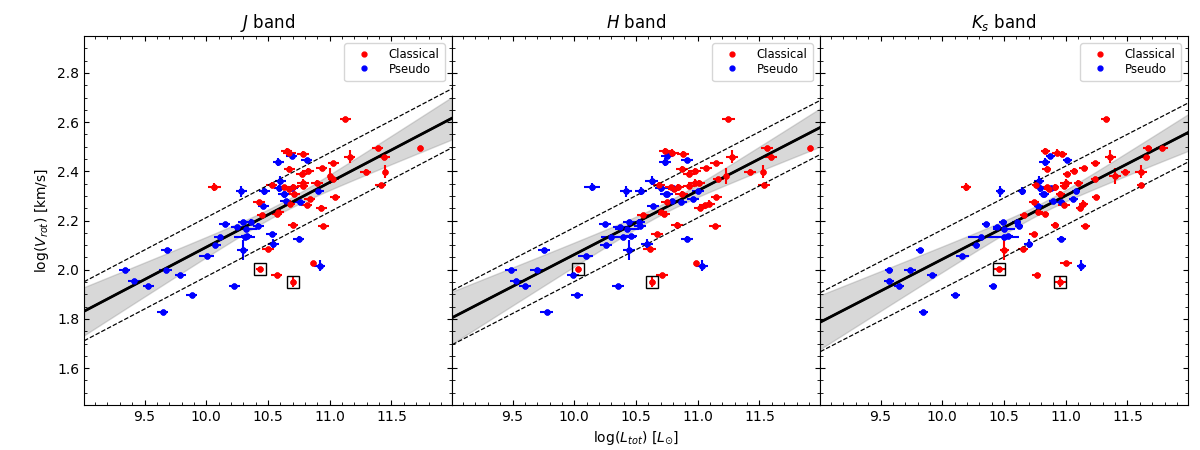}
\caption[TF]{\footnotesize The Tully-Fisher relation (TFR)  for the galaxies in our sample. Classical bulges and pseudobulges are coloured in the same way as previously. The three bands of 2MASS are shown. Symbols and colours are the same as in previous images. The black line is the linear fit for all disc galaxies. Black squares indicate the galaxies M~86 and NGC~2768 (see \S\ \ref{Dis_gal_scal_rel}).}
\label{TF}
\end{center}
\end{figure*}

\begin{figure*}
\begin{center}
\includegraphics[width=17.8 cm]{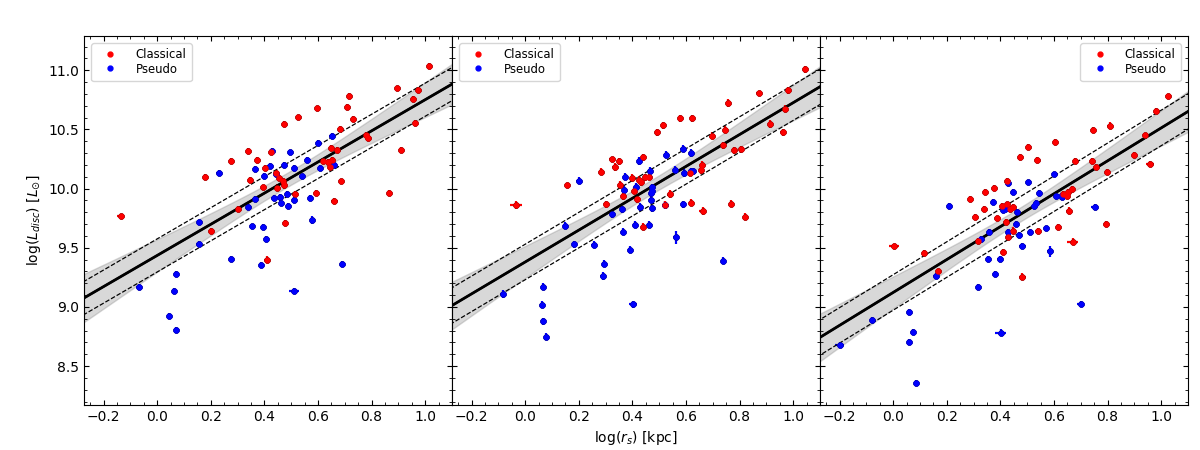}
\caption[L-rs]{\footnotesize The LSR$_d$ correlation between luminosity ($L_{disc}$) and scale length ($r_s$) for discs of galaxies in our sample. Discs of classical bulges and pseudobulges are coloured the same way as previously. The three bands of 2MASS are shown. Symbols and colours are the same as in previous images. The black line is the linear fit for all objects.}
\label{L_rs}
\end{center}
\end{figure*}

\subsubsection{Tully-Fisher (TFR) and Luminosity-Scale Length relations (LSR$_d$)} \label{TFR_S}

It has been found that the TFR has a dependence on galaxy morphology \citep[e.g.,][]{2011ARA&A..49..301V}, as well as the Luminosity-Scale Length (LSR$_d$) for discs, see \citetalias{Rios-Lopez_et_al_2021}.

To generate the TFR we use total absolute magnitudes for the spiral galaxies in \citetalias{Rios-Lopez_et_al_2021} and rotation velocities corrected for inclination ($V_{rot}$) from \texttt{HyperLeda}\footnote{\texttt{HyperLeda} reports an extreme value of $V_{rot}$ for the barred galaxy M~83. In this case. we corrected the rotation velocity following \citet{Courteau_et_al_2007} for inclination and redshift: $V_{rot}=\frac{V_{obs}}{\sin i (1+z)}$, where the inclination angle is given by  $i=\arccos\left(\sqrt{\frac{\left(\frac{b}{a}\right)^2-q_0^2}{1-q_0^2}}\right)$, where $\left(\frac{b}{a}\right)$ is disc's axis ratio from  \citetalias{Rios-Lopez_et_al_2021} and the flattening ratio of an edge-on spiral galaxy $q_0=0.2$ was held fixed.}. The ETG galaxies M~86 and NGC~2768 have reported $V_{rot}$ in \texttt{HyperLeda}; if we use the galaxies' absolute total magnitude, they become pronounced outliers to the TFR. However, if we take only the luminosity of the disc component, they fall closer to TFR, as shown in Figure \ref{TF}.

\subsection{SMBH Scaling Relations}
\label{SMBH_scal_rels}

We have explored known SRs for BH, considering bulges and pseudobulges. The information on BH masses limits our study; hence, our analysis is reduced to $\sim$ 1/3 of the galaxies in \citetalias{Rios-Lopez_et_al_2021}. Nevertheless, we still recovered relevant results, which agree with earlier works. The sample of pseudobulges with dynamically measured BH mass is small since we were left with only eight sources. The best fits resulting after a linear regression analysis are reported in Table \ref{Table_results_fits_M-rels} and shown in Figs. \ref{M-Lum} to \ref{M-V}.

Pseudobulges appear to follow the same SRs as bulges, i.e., in some cases, they do not present so much dispersion and populate the lowest part of the relations presented in this work. So, we have added more pseudobulges from the literature to explore whether or not these trends still prevail using larger bulge samples.  The additional $Ks$ band data were taken from \citet{Kormendy-Ho_2013} and \citet{deNicola_et_al_2019} and references therein. Empty blue circles represent the extra data in Fig. \ref{M-Lum} to \ref{M-V}, while parameters labelled with asterisks (*) mark the fits, including these data in Table \ref{Table_results_fits_M-rels}. We should remark that we did not include upper limits in the regressions; nevertheless, they are indicated as blue triangles in  Fig. \ref{M-Lum} to \ref{M-V} for comparison.

\subsubsection{Luminosity}

We have recovered the SRs between SMBH mass and bulge luminosity, $M_{\bullet}$-$L$, as well as with the total luminosity of the galaxy, $M_{\bullet}\texttt{-}L_{tot}$, shown in Figs. \ref{M-Lum} and \ref{M-Lum_tot}, respectively. After $M_{\bullet}$-$\sigma$ relation (see below), $M_{\bullet}\texttt{-}L$ is the one with less scatter, ranging from $\sim0.6-0.8\,{\rm dex}$, depending on the band and group of galaxies considered, while the \textbf{$M_{\bullet}\texttt{-}L_{tot}$} has a larger scatter, ranging from $\sim0.8-0.9\, {\rm dex}$. Pseudobulges follow the relation traced by SMBH masses and the luminosity for bulges, even when extra pseudobulges (empty circles in the right panel of Fig. \ref{M-Lum}) are added. At the same time, pseudobulges tend to be more dispersed when the total luminosity of the galaxy is considered, especially when the extra pseudobulges are added (right panel of Fig. \ref{M-Lum_tot}). We will discuss these trends in more detail in $\S$\ref{Dis}.

\begin{figure*}
\begin{center}
\includegraphics[width=18 cm]{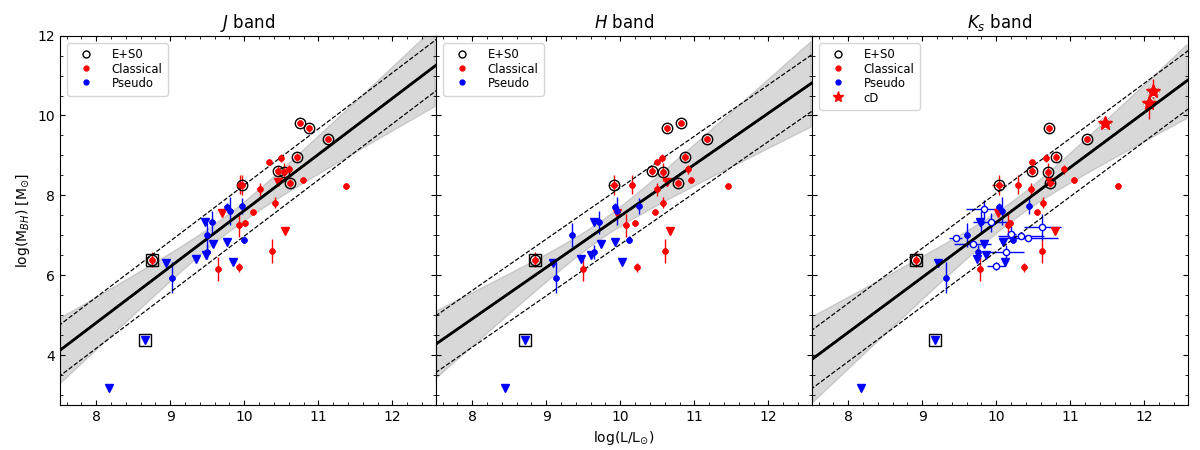}
\caption[M-Lum]{\footnotesize Correlation between the SMBH mass and luminosity of the bulge in the three bands of 2MASS. As previously mentioned, red-filled circles represent classical bulges, and the blue ones are pseudobulges according to our classification, while the red points inside black circles are the E+S0 galaxies. Symbols in triangles represent upper limits on SMBH mass. Empty blue circles represent pseudobulges taken from the literature (see \S\ \ref{SMBH_Masses_sample}). M~32 (classified as classical) and M~110 (classified as pseudo) are highlighted in black squares (see \S 5.1 in \citetalias{Rios-Lopez_et_al_2021}). Red stars represent the cD galaxies with measured BH mass. The grey-shaded region about the fit corresponds to the $1\sigma$ confidence band, while the dashed black lines indicate the intrinsic scatter of the fit for the entire sample.}
\label{M-Lum}
\end{center}
\end{figure*}

\begin{figure*}
\begin{center}
\includegraphics[width=18 cm]{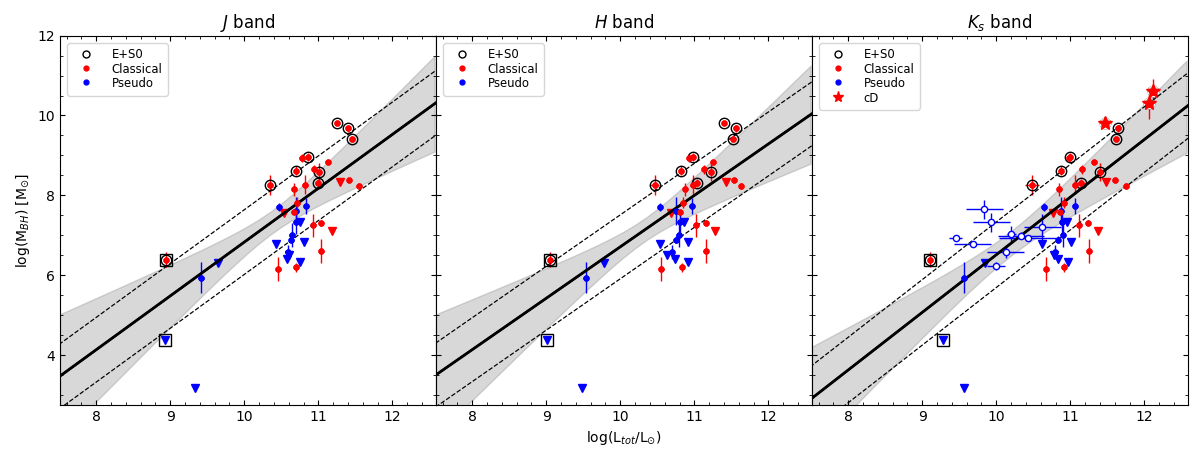}
\caption[M-Lum-tot]{\footnotesize Correlation between the SMBH mass and total galaxy luminosity in the three bands of 2MASS. Symbols and colours represent the same as in Fig. \ref{M-Lum}.}
\label{M-Lum_tot}
\end{center}
\end{figure*}

\subsubsection{Effective Radii}

\begin{figure*}
\begin{center}
\includegraphics[width=18 cm]{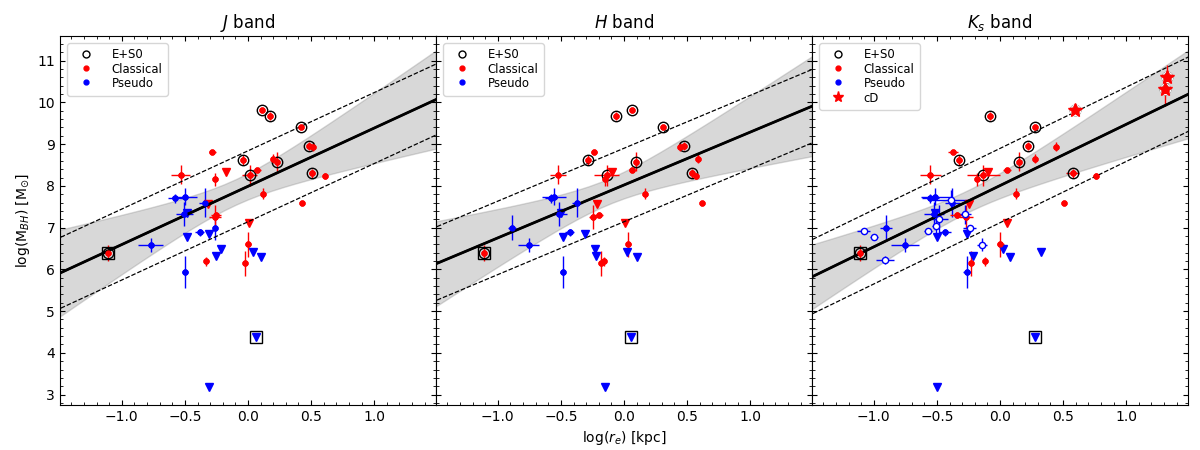}
\caption[M-re]{\footnotesize Correlation between the SMBH mass and effective radius in the three bands of 2MASS. Symbols and colours represent the same as in Fig. \ref{M-Lum}.}
\label{M-re}
\end{center}
\end{figure*}

Fig. \ref{M-re} shows the relation between BH mass and effective radius,  $M_{\bullet}$-$r_e$, showing a moderate correlation with a slope similar to $M_{\bullet}$-$L$ relation. However, its intrinsic scatter is larger than the scatter of the $M_{\bullet}$-$L$ relation but is very similar when compared to the one of $M_{\bullet}$-$L_{tot}$. Also, it is interesting to notice that for the $M_{\bullet}$-$r_e$, unlike other relations, pseudobulges tend to follow the same trend as the whole sample, even when the extra pseudobulges are added. Maybe it is just a coincidence, but this diagram's pseudobulge locus might not be random. A similar effect is seen in the $M_{\bullet}$-$V_{rot}$ relation.

\subsubsection{S{\'e}rsic Index ($n$)}

\begin{figure*}
\begin{center}
\includegraphics[width=18 cm]{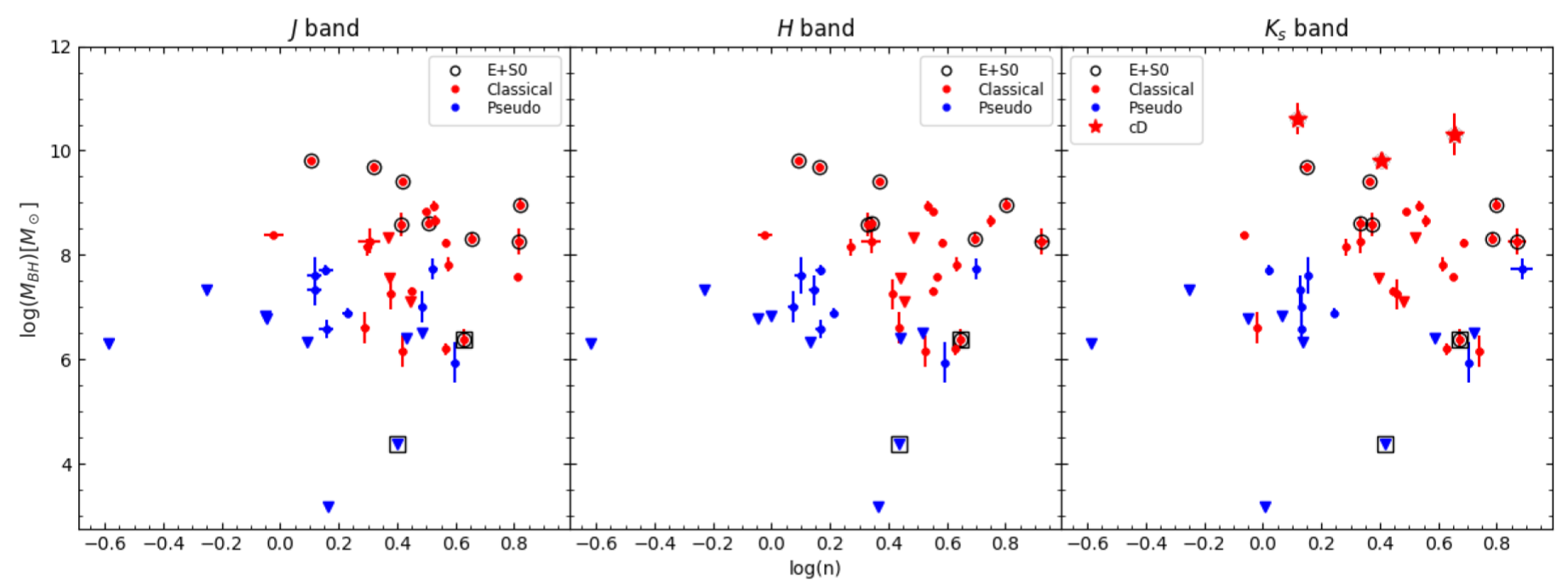}
\caption[M-n]{\footnotesize Correlation between the SMBH mass and S\'ersic index $n$ in the three bands of 2MASS. Symbols and colours represent the same as in Fig. \ref{M-Lum}.}
\label{M-n}
\end{center}
\end{figure*}

Some studies have suggested a connection between the mass of the SMBH and the S\'ersic index $M_{\bullet}\texttt{-}n$ \citep[e.g.,][]{Graham_2001, Graham_2007}. However,  as shown in Fig. \ref{M-n}, we do not find such a correlation. This correlation has the biggest dispersion of all the correlations involving SMBH mass studied here; furthermore, it has the lowest correlation coefficient (see Table \ref{Table_results_fits_M-rels}). This is contrary to \citet{Graham_2007}, who claims that the scatter of $M_{\bullet}\texttt{-}n$ relation is very similar to that of $M_{\bullet}\texttt{-}\sigma$.  Therefore, we conclude that $M_{\bullet}\texttt{-}n$ correlation is quite weak or insignificant in the NIR.  

\subsubsection{Velocity Dispersion}

The correlation between $M_{\bullet}$ and $\sigma$ and $\sigma_e$ are plotted in Fig. \ref{M-sig} (left and right panel, respectively). These relations present the lowest scatter among the SRs that we have considered in this study, with $\epsilon\sim0.5$ and $\sim0.3\, {\rm dex}$ for $M_{\bullet}\texttt{-}-\sigma$ and $M_{\bullet}\texttt{-}\sigma_e$, respectively. When more pseudobulges are considered, the $M_{\bullet}\texttt{-}\sigma_e$ relation tends to have a slightly larger scatter of $\sim0.4\, {\rm dex}$. These findings are in agreement with previous works, which point out that this relation tends to have a smaller scatter than the other SMBH scaling laws; in this way, $M_{\bullet}\texttt{-}\sigma_e$ is regarded as one of the best estimators of SMBH masses \citep[e.g.,][]{Kormendy-Ho_2013}.

\subsubsection{Rotation Velocity ($V_{rot}$) and Luminosity of the disc}

We also explore the relation outlined by the SMBH mass and the $V_{rot}$ of disc galaxies in Fig. \ref{M-V}. Our results indicate that this is a weak correlation. Also, the intrinsic scatter of $M_{\bullet}$-$V_{rot}$, as well as the slope, varies significantly for the complete sample in comparison to the classical and pseudo bulges subsamples: we found the scatter   $\sim0.9\texttt{-}0.3\, \rm{dex}$ for classical bulges and a scatter of  $\sim0.5\,\rm {dex}$ for pseudobulges, respectively. When more pseudobulges are considered, the trend is similar, and the scatter is slightly lower; besides, the correlation coefficients are higher in the pseudobulges subsamples. The latter one could be an indication that SMBH masses may show a distinct correlation with pseudobulges, or in a more general way, with late-type systems \citep[e.g.,][]{Beifiori_et_al_2012}.

We also investigated the link between SMBH mass and disc luminosity, and as can be seen in Fig. \ref{M-Disc}, we did not find a correlation for these parameters.

\subsection{Candidates for Dynamical Measurements of BH Masses}\label{Candidates}

Similarly, as \cite{Lopez-Cruz_et_al_2014}, in Table \ref{Table_candidates}, we present a list of attractive candidates for mapping the velocity fields on scales $\sim$ 0.40" (or even higher in some cases) through stellar- and gas-dynamical using space telescopes or the current instrumentation available at the largest ground-based telescopes, to perform a direct measurement of their BH masses.

The SMBH masses reported in Table \ref{Table_candidates} were estimated using SRs between $M_{\bullet}$ and bulge parameters, such as luminosity, velocity dispersion and effective radius. Such SR considered are those that we derived in this work, as well as additional relations reported in the literature, such as the ones from \citet{Kormendy-Ho_2013} and \citet{deNicola_et_al_2019}, to carry out a more robust estimate. Additionally, in the last column is shown the BH radius of influence, $r_f = GM_{\bullet}/\sigma^2$, where the SMBH significantly affects the stellar dynamics. This last parameter is reported in both angular and physical units.

\begin{figure*}
\begin{center}
\includegraphics[width=15.5 cm]{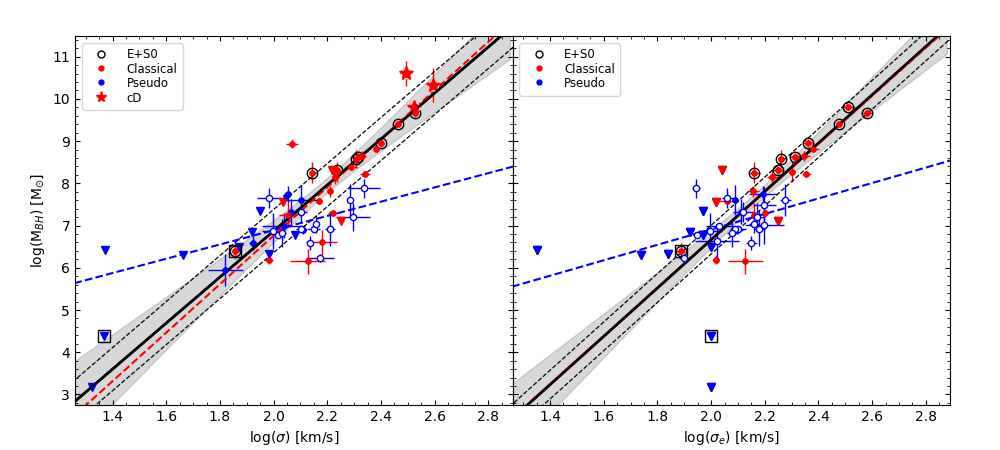}
    \caption[M-sig]{\footnotesize Left panel: Correlation between the SMBH mass and central velocity dispersion ($\sigma$). Right panel: Correlation between the SMBH and the effective velocity dispersion ($\sigma_e$) for some galaxies of our sample. Red and blue dotted lines are the linear fit for classical and pseudobulges. The solid black line fits the joint distributions of classical bulges and pseudobulges. Symbols and colours are the same as in Fig. \ref{M-Lum}.}
\label{M-sig}
\end{center}
\end{figure*}

\begin{figure}
%\begin{center}
\includegraphics[width=8.5 cm]{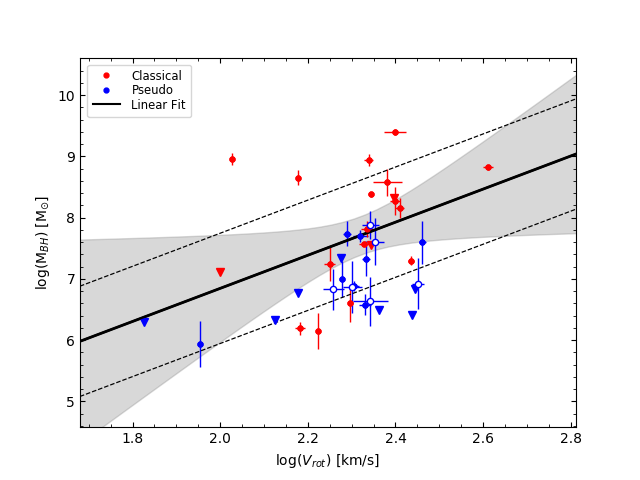}
\caption[M-V]{\footnotesize Relation between the SMBH mass and rotation velocity ($V_{rot}$) for some galaxies of our sample. Symbols and colours represent the same as in Fig. \ref{M-Lum}.}
\label{M-V}
%\end{center}
\end{figure}

\begin{figure}
\begin{center}
\includegraphics[width=8.5 cm]{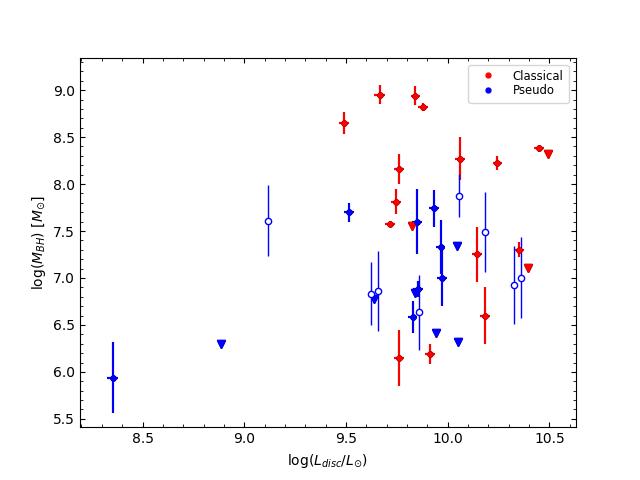}
\caption[M-disc]{\footnotesize Relation between SMBH mass and luminosity of discs for some galaxies of our sample. Symbols and colours represent the same as in Fig. \ref{M-Lum}.}
\label{M-Disc}
\end{center}
\end{figure}

\section{Discussion}\label{Dis}

\subsection{Galaxy Scaling Relations}\label{Dis_gal_scal_rel}

Our results on galaxy SR presented
above are in general agreement with those reported before, but after introducing the classification of bulges, we found deviations due to pseudobulges. Since we have covered a wide range of galaxy types, our results are pertinent to the evolution of both early and late-type galaxies.

The LCRs shown in  Fig. \ref{n-AbsMag} tell us that brighter bulges and galaxies have higher concentration indexes ($n$) and vice versa. In that sense, our result is consistent with previous studies \citep[e.g.,][]{Graham_2001, Fisher-Drory_2008, Anorve_PhD_Thesis, Graham_2013}. Pushing this trend, some authors have suggested that $n$ is physically meaningful \citep[e.g.,][]{Blanton_et_al_2003, Anorve_PhD_Thesis, Kormendy_et_al_2009}; nevertheless, this relation has the widest spread around its fits, resulting in a weak correlation, as seen in most cases (see Table \ref{Table_results_fits_FP_projections}). However, we highlight that the scatter of the LCR is less significant for cD galaxies alone; it runs almost parallel but with a flatter slope above the LCR of classical bulges. Furthermore, when considering the entire sample, including cD galaxies, the LCR's slope is steeper. However, as indicated in Figure \ref{n-AbsMag} and Table \ref{Table_results_fits_FP_projections}, we conclude that the LCR for the whole sample and its subsamples have a large scatter. We can only declare that more luminous galaxies tend to have higher Sérsic indexes, $n$; however, cDs do not reach large $n$ as we would expect from the LCR for bulges but become clear outliers, Figure \ref{n-AbsMag} also suggests that galaxies with $n>10$ are unexpected or at least rare. Nevertheless,  cD galaxies, on a first approximation, seem to fall on the FJR, the LSR and the FP as extensions of classical bulges, as shown in Figures \ref{FJR_final}, \ref{LSR_final}, and \ref{FP_final}, respectively. 

We can see how pseudobulges populate the lower part of the correlation, while classical bulges occupy the higher regions in the FJR shown in Fig. \ref{FJR_final}. This trend agrees with previous results \citep[e.g.,][]{Kormendy-Kennicutt_2004, Jiang_et_al_2011, Anorve_PhD_Thesis}, confirming the idea that the FJR can help segregate classical bulges from pseudobulges \citep[e.g.,][]{Anorve_PhD_Thesis,Kormendy-Ho_2013}, showing that pseudobluges have low-$\sigma$ in the FJR. This behaviour might result because these systems are rotationally supported; therefore, their velocity dispersion is lower. On the other hand, ellipticals and classical bulges are formed via major mergers \citep[e.g.][unlike pseudobulges, which are shaped mainly through secular processes]{Kormendy-Kennicutt_2004,2013seg..book....1K, 2013pss6.book..265B} and, as a result, random motions support their structure and thus with higher values of the velocity dispersion. In addition, the cD galaxies seem an extension of the FJR for classical bulges; however, cD galaxies alone deviate from having a flatter slope but similar dispersion. 

The recent study of \citet{2024MNRAS.527..249Q} for massive ETGs shows that the less luminous galaxies in the ATLAS sample follow a steeper FJR. Although our sample includes more spiral galaxies than ETG, we suggest that the change reported by 
\citeauthor{2024MNRAS.527..249Q} is due to the inclusion of pseudobulbes ($\logd \sigma < 2.1$), as seen in our Figure \ref{FJR_final}.

For the LSR in Fig. \ref{LSR_final}, we obtained similar results to those reported in the literature \citep[e.g.,][]{Nair_et_al_2010, Cappellari_et_al_2011, Anorve_PhD_Thesis}. We noticed, however,  that  \citet{Nair_et_al_2010} used $r_{p90}$, the Petrosian radius containing 90\% of total light, instead of the effective radius as in this work, but the results are alike. Nevertheless, when the LSR is generated with the $r_{p90}$, the dispersion is lower than that derived using $r_e$ \citep{Nair_et_al_2010, Nair_et_al_2011}. We agreed with previous studies that pseudobulges do not follow the same relation as classical bulges. Most pseudobulges fall below classical bulges in these relations, showing a wider scatter than in prior cases. Likewise, our slope and intercept are consistent with those from \citet{Lasker_et_al_2014}, who also constructed the LSR with data in the $K_s$ band using a 2D photometric decomposition. cD galaxies follow the LSR due to classical bulges with similar dispersion; however, the dispersion of cD galaxies' LSR is smaller than the one of pseudobulges and the total samples (both \textit{All} and \textit{All+cD}). From the comparison of the LSRs in our Figure \ref{LSR_final} and Figure 3 in \citet{2024MNRAS.527..249Q}, we are suggesting that the LSR slope change in \citet{2024MNRAS.527..249Q}, is due to pseudobulges. Nevertheless, our suggestion needs further confirmation using a larger sample of ETG. 

Concerning the FP presented in Fig. \ref{FP_final}, highlighting classical bulges and pseudobulges as in previous images, the expressions obtained are consistent with those derived in NIR bands by \citet{Pahre_et_al_1998}. In the correlations studied so far, pseudobulges tend to populate the lower parts of the diagrams, showing larger dispersions than classical bulges and cD galaxies located in the higher parts of them. Nevertheless, on the FP, the separation between bulges and pseudobulges is less apparent \citep[c.f.,][]{2024MNRAS.527..249Q}, while the dispersion of pseudobulges about the FP is slightly larger than that of classical bulges. Regarding cD galaxies, these also show a large dispersion about the FP, which might indicate that additional processes are occurring throughout the formation of supergiant galaxies.

The CMR (\S \ref{Col-Mag}) for bulges and discs in Fig. \ref{CMR} and \ref{CMR_discs}, respectively, we found that the colours of classical bulges and pseudobulges are constant inside a nine-magnitude range ($-26 \leq M_{K_s} \leq -17$). Previous studies in the optical or the NIR  \citep[e.g.,][]{Lopez-Cruz_et_al_2004} used colour apertures to generate the CMR for cluster galaxies and, in some cases, field ETGs, finding negative slopes. The galaxies in our sample are mostly spiral drawn from low-to-high density environments \citep[cf.,][]{Jarrett_2004}; nevertheless, the shallow slope ($m=-0.025 \pm 0.009$) in the $J-K\, \mathrm{vs}\,H$ diagram reported by \citet{2007ApJS..169..225E} for galaxies in the Coma cluster and for ETG reported by \citet{2016AJ....152..214S} ($m\sim 0.021$, $J-K \, \mathrm{vs}\, M_J$), agree with the CMR that we found for discs ($m_{disc}=-0.022 \pm 0.014$), see Table \ref{Table_results_fits_FP_projections}. However, the intercepts should differ since we use disc colour and luminosity (absolute magnitudes).  

The lack of correlation between bulges and pseudobulges colours and luminosity in the NIR colours might suggest that we are tracing the bulge's old stellar population (about 10 Gyr old). For this component, metallicity contributions cannot be traced using broad-band NIR colours \citep[e.g.,][]{2009MNRAS.392..982C}. Indeed, the studies of \citet{1997A&A...320...41K} and \citet{2001ApJ...551L.127V} suggested for old stellar populations, age-metallicity indicators are found around $4000\,\mathrm{\AA}$ break and the Mg complex.  Indeed, Fig. \ref{CMR} shows that classical bulges tend to be more luminous than pseudobulges, and there is an evident division between classical bulges and pseudobulges. Using this result, we can form a criterion to distinguish between bulge types using the division indicated by the vertical dashed line at $M_{K_s}=-22\;\mathrm{mag}$ in Fig. \ref{CMR}. Hereafter, classical bulges can also be defined by their luminosity: $M_{K_s}\leq-22\;\mathrm{mag}$; however, if it's less luminous than this threshold, it's a pseudobulge. A similar criterion was proposed by \citet{Fisher-Drory_2016} in the optical, in which pseudobulges were bluer than $B-V\sim0.5$, while classical bulges tend to be redder ($B-V\gtrsim 0.5$).

The large dispersion in the CMR for classical bulges might be due to selection effects; for instance, we highlighted in Fig. \ref{CMR} the E+S0 galaxies that are hosting an AGN or have peculiar classification \citepalias[the latter one is taken from 2MASS, see Table 1 in][]{Rios-Lopez_et_al_2021}. Although we have modelled the AGN contribution as a point source, we suppose that  AGN extended light might be increasing the dispersion to the mean constant NIR colour of bulges ($\mean{J-H} = 0.7$, $\mean{H-K_s}=0.3$, and $\mean{J-K_s}=1.0$), as can be seen for the galaxies  M~60, NGC~1549, NGC~4976, NGC~5084 and NGC~2768 \citep[][]{Shurkin_et_al_2008, Diehl_et_al_2007,deVaucouleurs_1964,deVaucouleurs_et_al_1991,Irwin_2019}, which appear as the most extreme outliers (highlighted in green squares) in  Fig. \ref{CMR}.

We found that discs in galaxies with classical bulges tend to be redder than those with pseudobulges. We suggest that the slope of the  CMR  for discs is due to the spread in age for star formation activity and its dependence on radius in galaxies' discs \citep[e.g.,][]{2014A&A...562A..47G}. Since the mass of the disc is  $M_{disc}\propto r\times V_{rot}^2$ by the TFR and the LSR$_d$ (using values from Table \ref{Table_results_fits_FP_projections}), we find $M_{disc} \propto L^{\frac{4}{3}}$; therefore, from the CMR for disc we have that more massive discs are redder in the NIR \citep[cf.,][]{2001ApJ...550..212B}.

Moreover, the slight gradient found in the disc colours may be related to specific physical mechanisms within the galaxy, such as the inside-out formation scenario, where the central regions of galaxy discs formed first and are less rich in gas and show a lower sSFR than the outer regions  \citep[e.g.,][]{Gonzalez_Delgado_2015, Lian_2017, Ellison_2018}.
For instance, more massive galaxies (corresponding to the most luminous ones) have experienced a more efficient quenching process of their star formation than the less massive ones (the less luminous ones), turning a galaxy red. Similarly, more massive discs have quenched their star formation more efficiently than the less massive discs \citep[e.g.,][]{2017_CALIFA_SF}. Also, \citet{2018_Belfiore} reported that the specific star formation rate (sSFR) does depend on the mass in regions dominated by the galaxy disc since high-mass galaxies show lower sSFR than the low-mass ones at large radii ($r > 1.5~r_e$) corresponding to the disc component, which agrees with the inside-out growth scenario, where central regions of galaxy discs are formed earlier, implying that are more evolved, less gas-rich and have lower sSFR than the outer regions of the disc.

Figure \ref{CMR_discs} is in close agreement with Coma cluster galaxies' CMR  in the NIR  reported by  \citet{2007ApJS..169..225E} (see their Table 13 and  Figure 3), we conclude that the slope of the CMR reported by \citeauthor{2007ApJS..169..225E} and \citet{2016AJ....152..214S} is due to the colours of discs rather than the colours of bulges; This distinction was previously overlooked, likely because colour-apertures are unable to differentiate among the colours of different morphological components.

The TFR's slopes found in this paper (see Table \ref{Table_results_fits_FP_projections} and Fig. \ref{TF})  are in excellent agreement with the study of  \citet{Courteau_et_al_2007} who performed a comprehensive analysis for a large sample of spiral galaxies ($\sim$1300 sources) in the $I$ and $K_s$ bands. They reported slopes in the $K_s$ of 2MASS from 0.25 to 0.27 depending on the (sub)sample considered, from Sa to Sd Hubble types and for the whole sample, while our slopes are in the same range of values in the three NIR bands for the entire set of galaxies. Similarly, \citet{Courteau_et_al_2007} also constructed the LSR for galaxy discs with a slope of 1.76 for the Sa galaxies of their sample, which is in agreement considering errors with ours of 1.39, as well as in the scatter ($\sim$0.27) of the relation (see Table \ref{Table_results_fits_FP_projections} and Fig. \ref{L_rs}). Furthermore, in our considerations for galaxy discs, the previous trends for bulge types still prevail because the separation between classical and pseudos is noticeable. In addition, we note that the slopes of both luminosity-size relations drawn in this work are very close, i.e., a value $\sim$1 for the LSR, while for the LSR$_d$ is slightly steeper $\sim$1.3. The above results are for the entire sample, while for classical bulges ($\sim$0.95) is quite similar; nevertheless, it is interesting that the slopes of the LSR for pseudobulges are $\sim$0.20. This might indicate that pseudobulges do not have physical properties similar to galaxy discs. Still, they represent a complex transient morphological feature related to dynamical mechanisms similar to galactic bars and other structural components (such as oval discs, inner and outer rings, among others) present in galaxy discs \citep[e.g.,][]{Bournaud_et_al_2005, 2013seg..book....1K}.

We have used the luminosity of the disc component to place the ETGs M~86 and NGC~2768 closer to TFR, indicated by red dots (classical bulges) enclosed by a black square in  Fig. \ref{TF}.  \citet{2007ApJ...668...94H} used 2MASS $K_s$ photometry and velocities from  \texttt{HyperLeda} finding a large number of outliers to the TFR across the Hubble sequence, and singled out galaxies too bright for their $V_{rot}$. We propose that, at least for ETG, using the luminosity of the disc gives a better approximation to the TFR because embedded discs are reacting to the mass enclosed by a sphere or cylinder of the same radius (bulge stars and dark matter). We also note that bars can affect disc inclination measurements, such as M~83; see footone in \S \ref{TFR_S}. \citet{2007ApJ...668...94H} put attention to the inclination determination by resorting to published axes ratios to calculate the disc inclination by applying a formalism similar to ours for M~83;  therefore, we suggest that we may reduce the scatter of the TFR in Figure \ref{TF} and severely reduce the number of outliers if the inclination is calculated using axes ratios and the luminosity of embedded discs from a detailed SB analysis as the one presented in \citetalias{Rios-Lopez_et_al_2021} \citep[see][for example]{2013AJ....146...86T}. From Figure \ref{TF}, we suggest that classical bulges are preferentially found in luminous spiral galaxies with fast rotating discs ($V_{rot}\gtrsim \mathrm{160 \; km\,s^{-1}}$) 

\subsection{SMBH Mass Scaling Relations}

We show the correlation between $M_{\bullet}$ and bulge luminosity shown in Fig \ref{M-Lum}. 
Many studies have claimed that the $M_{\bullet}\texttt{-}L$ relation shows a similar scatter as the $M_{\bullet}\texttt{-}\sigma$. In our case, the $M_{\bullet}\texttt{-}L$ correlation, after $M_{\bullet}\texttt{-}\sigma$, has a smaller scatter than the other correlations listed in Table \ref{Table_results_fits_M-rels}. These trends agree with previous works of bulges in the NIR. Besides, our values for the slopes are consistent within uncertainties, although our scatter coefficients are slightly bigger \citep[e.g.,][]{Marconi-Hunt_2003, Kormendy-Ho_2013, deNicola_et_al_2019}. It found that Holm~15A, NGC~4889 and M~87 follow the general $M_{\bullet}\texttt{-}L$ in the $K_s$ band \citep[cf.,][]{Mehrgan_et_al_2019}. Furthermore, the relation for the bulge component shows less dispersion than the one for the total luminosity of the galaxy (see Fig. \ref{M-Lum_tot}), consistent with the result of \citet{Lasker_et_al_2014-2}, who reported scatters of $0.48\,\rm{dex}$ and $0.51\,\rm{dex}$, for bulge and total luminosity in $K$ band, respectively; while our scatters are 0.68 and 0.83, respectively, in $K_s$ band. For classical and pseudo bulges, \citet{deNicola_et_al_2019} reported slopes of $1.00 \pm 0.09$ and $0.49 \pm 0.47$, respectively, while ours are $1.33 \pm 0.24$ and $2.08 \pm 0.44$; however, when more pseudobulges from literature are added, the slope turns out to be $0.30 \pm 0.31$, very close to the one of \citet{deNicola_et_al_2019}. Hence, so far, we can conclude that pseudobulges follow a different $M_{\bullet}\texttt{-}L$ (see \S\ref{behave_pseudos} for a further discussion on this issue).

Regarding the $M_{\bullet}\texttt{-}r_e$ correlation presented in Fig. \ref{M-re}, it has very similar slopes in the three NIR bands as the $M_{\bullet}\texttt{-}L$ and $M_{\bullet}\texttt{-}L_{tot}$, although the latter shows less scatter than the former one. These trends coincide with the findings of \citet{deNicola_et_al_2019}. Also, the slopes in $K_s$ for the whole sample and classical bulges are very similar to those reported by \citet{deNicola_et_al_2019}, but ours are more dispersed; similarly, in the case of pseudobulges, where ours have higher dispersion. Nevertheless, the slope for pseudobulges is closer to the one of \citet{deNicola_et_al_2019}, and the intrinsic scatter is reduced when extra pseudobulges are included (see Table \ref{Table_results_fits_M-rels}). Interestingly, we could expect a tighter correlation between BH mass and the galaxy's size because of the FP and their projections, where $r_e$ clearly shows a connection with luminosity and velocity dispersion. Nonetheless, as we said before, we found a moderate correlation ($r\sim 0.5$), albeit \citet{Fabian-Lasenby_2019} argue that the effect of the SMBH on its host galaxy would be detectable since the energy released during the AGN phase by the BH may have an essential role in the final stellar mass, as well as likely in the size of the galaxy.

As we already indicated, the relation between $M_{\bullet}$ and $n$ that we presented here is weak (see Fig. \ref{M-n} and Table \ref{Table_results_fits_M-rels}), which is in agreement with previous studies that also find a poor correlation \citep[e.g.,][]{Beifiori_et_al_2012, Kormendy-Ho_2013}. The weakness of this correlation can be expected since we also found a weak LCR (see Fig. \ref{n-AbsMag} and Table \ref{Table_results_fits_FP_projections}). Additionally, this correlation has the highest dispersion and the lowest correlation coefficient ($>1 \,dex$ and $r\sim 0.20$ or even lower, respectively, for most cases; see Table \ref{Table_results_fits_M-rels}), even more than the relations involving disc parameters. Therefore, the fact that the $M_{\bullet}\texttt{-}n$ correlation is noisier may indicate that it is not as fundamental as the other SRs involving velocity dispersion, luminosity, or even the effective radius. Therefore, the lack of correlation  $M_{\bullet}\texttt{-}n$ implies that $n$ is not an SMBH mass estimator. 

We found that the $M_{\bullet}\texttt{-}\sigma_e$ relation (right panel of Fig. \ref{M-sig}) has the lowest intrinsic scatter out of all the relations considered in this study. This finding agrees with previous works \citep[e.g.,][]{Ferrarese-Merritt_2000, Marconi-Hunt_2003, Kormendy-Ho_2013, Saglia_et_al_2016, 2016ApJ...817...21S, deNicola_et_al_2019}. Also, the intrinsic scatter and other parameters of the relations between $M_{\bullet}$ with $\sigma_e$ and $\sigma$ (left panel of Fig. \ref{M-sig}) differ just slightly, being smaller the one of $M_{\bullet}\texttt{-}\sigma_e$, even when more pseudobulges are added as reported in Table \ref{Table_results_fits_M-rels}. This can be expected since, as \citet{Kormendy-Ho_2013} indicated, the difference between central and effective velocity dispersions is minimal (they also used $\sigma$ from \texttt{HyperLeda} as in our case). Then, the effect on the scatter for $M_{\bullet}$ with these two parameters is insignificant. Besides, our slope coefficients are compatible (within the errors) with previous works; for instance, for $M_{\bullet}\texttt{-}\sigma$, we report $5.45 \pm 0.52$ using the complete sample, while \citet{Ferrarese-Merritt_2000} reported a value of $4.80 \pm 0.54$ and \citet{Kormendy-Ho_2013} a slope of $4.38 \pm 0.29$. It has been suggested that the $M_{\bullet}\texttt{-}\sigma$ relation breaks down for  galaxies with $\sigma \geq \mathrm{270\; km\, s^{-1}}$ \citep[e.g.,][]{Lauer_et_al_2007,Kormendy-Ho_2013}; however, Figures \ref{M-sig} shows no signs of a breakdown at $\logd \sigma \geq 2.43$ and  Holm~15A presents only a slight deviation, in agreement with \citet{Mehrgan_et_al_2019}. In addition, aperture corrections for velocity dispersion measurements are based on velocity dispersion profiles typically described as a power-law function, where it has been demonstrated that there is a variation in the power-law slope \citep[e.g.,][and references therein]{Zhu_et_al_2023}. For instance, velocity dispersion values from \texttt{HyperLeda} are corrected according to the reported power-law slope of -0.04 by \citet{Jorgensen_et_al_1995} for early-type galaxies. However, using integral field unit (IFU) spectroscopy in a sample of galaxies not selected by morphology, \citet{deGraaff_et_al_2021} found a slope value of -0.033 $\pm$ 0.003, which is very close to the \cite{Jorgensen_et_al_1995} value.

For the $M_{\bullet}\texttt{-}\sigma_e$ relation, our slope values are $5.72 \pm 0.49,\, 5.69 \pm 0.55,\,\rm{and}\, 2.90 \pm 3.50$ for the whole, classical and pseudobulge samples, respectively. In comparison with \citet{deNicola_et_al_2019}, who also divided bulge types, they reported values of $5.07 \pm 0.27,\,4.48 \pm 0.30\, \rm{and}\, 3.50 \pm0.70$ for the same groups of objects. When more pseudobulges with $\sigma_e$ and $\sigma$ values are added, the slopes for the entire and pseudobulge samples decrease (see Fig. \ref{M-sig} and Table \ref{Table_results_fits_M-rels}). Again, we can conclude that classical bulges and pseudobulges follow very different $M_{\bullet}\texttt{-}\sigma$ relations, regardless of the radius where $\sigma$ is measured.

The relations of $M_{\bullet}$ with disc parameters such as  $V_{rot}$ and disc luminosity $L_{disc}$, \citet{Beifiori_et_al_2012} found a poor connection for $M_{\bullet}\texttt{-}V$, while \cite{Kormendy-Ho_2013}, more generally, pointed out that SMBH masses do not correlate with a disc of galaxies because they did not find such correspondence in $M_{\bullet}\texttt{-}L_{disc}$. Figures \ref{M-V} and \ref{M-Disc} and Table \ref{Table_results_fits_M-rels} confirm such trends. \cite{Kormendy-Ho_2013} still go further by arguing that because of the contribution of the disc, the  $M_{\bullet}\texttt{-}L$ is lost if the total luminosity of the galaxy is instead of the luminosity of the 
bulge. Nevertheless, as \cite{Lasker_et_al_2014-2} said, the galaxy's total luminosity can also be a reliable indicator for SMBH-galaxy coevolution since this relation can be characterised more robustly.  Figure \ref{M-Lum_tot} indicates \textbf{$M_{\bullet}\texttt{-}L_{tot}$} appreciably weakens. 

Therefore, our results suggest that SMBH masses are more closely linked to velocity dispersion and luminosity than other galaxy parameters, mainly with the former one, which in turn indicates that $\sigma$ \citep[with its caveats, e.g.,][]{Lauer_et_al_2007}, can be considered as one of the best estimators of $M_{\bullet}$. Additionally, our findings confirm previous results about the SMBH coevolution with elliptical galaxies, mostly with classical bulges rather than pseudobulges. Nevertheless, as we indicate below in \S\ref{behave_pseudos}, paying attention to the trend displayed by pseudobulges in BH relations is worthwhile.

\subsection{A general perspective on the BH scaling laws interpretation.}

The differences in the characterisation of our SRs for black holes and those reported in the literature may be related to sample size and the type of galaxies considered, which might involve the galaxies' structure. 

In this paper, we have explored the SRs for SMBHs for 33 galaxies, which are approximately a factor of 3 smaller than the samples considered by \cite{Kormendy-Ho_2013, Lasker_et_al_2014-2, Saglia_et_al_2016, deNicola_et_al_2019}. These samples were dominated by E+S0 galaxies and considered bulge/pseudobulge classifications. Moreover, \cite{Saglia_et_al_2016} pointed out the fact that the scatter for BH correlations tends to be smaller using only cored elliptical galaxies, agreeing with  \cite{Peng_2007}, who highlighted that mergers play a critical role in decreasing the scatter (below we will return to this point). Similarly, when barred galaxies are removed,  the $M_{\bullet}\texttt{-}\sigma$ slope decreases to 3.68 \citep{Graham_2008}. \cite{Lasker_et_al_2014-2} also used a sample slightly more extensive than ours, focusing on ETGs with very few spiral galaxies (only 4). In the same way,  our galaxy sample is marginally smaller than the one compiled by \cite{Marconi-Hunt_2003}, although their sample has only a quarter of spiral galaxies. In our sample, slightly more than two-thirds are spiral galaxies (generally, we still cover from elliptical to barred spiral galaxies).

 Slope variations might arise mainly from the differences in the velocity dispersion measurements \citep{Tremaine_et_al_2002} and the photometric decomposition to measure the luminosity. We agree with \cite{Lasker_et_al_2014-2}, who suggested that NIR photometry and 2D photometric analysis can provide more robust measurements of the bulge and the total galaxy luminosities. Thus, \cite{Gebhardt_et_al_2000} reported a slope of $3.75 \pm 0.3$, while \cite{Ferrarese-Merritt_2000} values of $4.80 \pm 0.5$; further works introduced refinements, reporting  slopes of $4.65 \pm 0.42$ \citep{Merritt-Ferrarese_2001}, $4.02 \pm 0.32$ \citep{Tremaine_et_al_2002} and $4.86 \pm 0.43$ \cite{Ferrarese-Ford_2005}. As shown in Table \ref{Table_results_fits_M-rels}, our slopes are $\sim5$ (however, for pseudobulges is smaller, $\sim$ 3). Meanwhile, the intrinsic scatters we report tend to be slightly larger, mainly for the correlations of $M_{\bullet}$ with luminosity and effective radius.

Additionally, we address the interpretation of SRs in a general theoretical context. It is still puzzling the existence of these correlations, as well as the fact that some of them present very small scatter, which has led to a common perspective that the growth of BHs is closely connected to the evolution of their host galaxies \citep[e.g.,][]{Kormendy-Ho_2013}. A possible explanation for empirical relations, such as $M_{\bullet}\texttt{-}M_{Bul}$ and $M_{\bullet}\texttt{-}\sigma$, is related gas accretion that allows SMBHs growth. This process will also produce the stellar bulge's growth with the BH. In this scenario, feedback from the AGN plays an essential role since the energy produced by the BH blows away the gas, causes the quenching of star formation and finishes the BH growth itself by halting the gas accretion \citep[e.g.,][]{Silk-Rees_1998, DiMatteo_et_al_2005, 2018MNRAS.479.4056W}.

We suggest that the slope of $M_{\bullet}\texttt{-}\sigma$ may also have theoretical implications related to this AGN feedback mechanism, on whether the feedback is mainly via energy or momentum transfer \citep[e.g.,][]{Silk-Rees_1998, King-Pounds_2015}. As we can see, our findings suggest a relation of the form $M_{\bullet} \propto \sigma^5$, corresponding to energy-driven winds \citep[e.g.,][]{Silk-Rees_1998}. On the other hand, momentum-driven winds result in a relation where $M_{\bullet}$ scales with $\sigma^4$ \citep[e.g.,][]{King-Pounds_2015}.

From a simple derivation, BH masses are related to bulge masses by assuming a fairly constant ratio between $M_{\bullet}$ and bulge mass, $M_{Bul}$ \citep{Haring-Rix_2004}. Due to the empirical relation $M_{\bullet} \propto \sigma^{\xi}$, we also have that $M_{Bul} \propto \sigma^{\xi}$. From the FJR, we know that the luminosity of the bulge scales with $\sigma^{4}$ \citep{Faber_et_al_1987}, then we find $M_{Bul}\propto\sigma^{5}$. Hence, $M_{\bullet} \propto \sigma^{5}$ agrees with our findings and the AGN feedback of energy-driven winds. In a similar fashion, $M_{\bullet} \propto L^{1.25}$, which shows consistency with our slopes (see Table \ref{Table_results_fits_M-rels}) not only with bulge luminosity but also for the total luminosity of galaxy (except for pseudobulges).

No cD galaxy has been found in the field; then cD galaxies might formed by different processes \citep[e.g.,][]{1997ApJ...475L..97L,2020ApJS..247...43K}, which are particular to clusters of galaxies \citep[e.g.,][]{2021MNRAS.507.4016D}.  Therefore,  the processes that help SMBH growth are possibly associated with cD galaxy growth. Mechanisms such as wet and dry mergers,  BH mergers, in situ and ex situ star formation, gravitational wave recoils, compact star cluster formation, dynamical friction, cusp scouring, and cooling cores, to name a few, are relevant at a particular epoch during SMBH-cD galaxy coevolution \citep[e.g.,][]{1990dig..book..394T,1996Natur.379..613M,1997ApJ...475L..97L, 2015ApJ...807...56B, 2018ApJ...864..113R,2018MNRAS.479.4056W,2023MNRAS.521..800M}.

\cite{Peng_2007} presented an alternative explanation for the origin of the SMBH scaling relations, mainly from a statistical perspective. In this scenario, mergers will generate these correlations as a purely stochastic process given by the central limit theorem, for instance, a sequence of a few random major mergers of galaxies containing seeded BHs. On the other hand, \citeauthor{Peng_2007} argues that AGN feedback is not the dominant process because some simulations reproduced SMBH scaling relations without feedback.  Another relevant issue is that major mergers occur mainly in the high regime. In contrast, minor mergers contribute to having a linear connection; the tendency to have a smaller scatter is weaker in the lower part of the relation. As observational evidence indicates, the latter argument can also be associated with pseudo-bulges populating the lower parts of BH correlations. Although the primary mechanism for pseudobulge growth is mainly via secular processes, it is also expected that these objects have passed through a quiet merger history and flyby interactions \citep[e.g.,][]{Izquierdo_2019, 2013pss6.book..265B, Kormendy-Kennicutt_2004, Kumar_et_al_2021}.

\subsection{The behaviour of pseudobulges in BH relations} \label{behave_pseudos}

In the SMBH scaling relations from Figs. \ref{M-Lum} to \ref{M-sig}, pseudobulges populate their lower parts (even for weak LCR). Furthermore, in the FJR, the LCR, the TFR and the CMR pseudobulges segregate in a way that tends to occupy the low-luminosity regimes of the correlations, which hints at a similar response observed in the SMBH scaling relations. Similar trends for pseudobulges in the BH scaling laws have been reported before, specifically for the relation between $M_{\bullet}$ and the bulge mass \citep[e.g.,][]{Greene_et_al_2008, Sahu_et_al_2019_2}. \cite{Saglia_et_al_2016} reported a weak correlation for SMBHs and pseudobulges only for the $M_{\bullet}\texttt{-}\sigma$. \cite{Sani_et_al_2011}, using a reduced sample of 9 pseudobulges (classified only with the S\'ersic index criterion), also reported that they tend to fall within the scatter of the BH relations involving luminosity, size and velocity dispersion. Similarly, \citet{Bennert_et_al_2021}, using a sample of 21 pseudobulges (identified with a robust scheme; also 26 barred galaxies, which could probably host pseudobulges) suggested that pseudobulges, barred spirals and objects with signs of merger activity follow the BH correlations. This, in turn, may be important because the secular evolution mechanism would be more relevant for the growth of SMBHs. The implications would further explain the origin of SMBH's SR since the galaxy merger process may not be regarded as the main channel for the coevolution of BHs and their hosts.

Before claiming that such a trend is valid, we should point out that the size of our pseudobulge sample with detected BH masses is limited (that is why we did the exercise of including more pseudobulges with complementary data from the literature; see plots for the $K_s$ band in Figs. from \ref{M-Lum} to \ref{M-sig}). Nevertheless, considering pseudobulges alone, we can establish that classical bulges and pseudobulges follow different SR, as our results indicate from their distribution in the plots and the linear regression analysis. Hence, instead of suggesting that pseudobulges follow the same correlations between $M_{\bullet}$ and galaxy parameters traced by ellipticals and classical bulges,  our results indicate a weak correlation in most cases, primarily due to the scarcity of pseudobulges in our sample (and even for some of the works mentioned above).

Furthermore, in \citetalias{Rios-Lopez_et_al_2021}, we ignored composite systems, where the coexistence of both pseudobulges and classical bulges might be present \citep{Erwin_et_al_2015, Fisher-Drory_2016}. Since classical bulges correlate better with the $M_{\bullet}$, we can not discard the presence or not of such composite systems in those pseudobulges lying in the relations for SMBHs.

Additionally, in Table \ref{Table_results_fits_M-rels} and Fig. \ref{M-sig}, a shallower slope for the subsamples of pseudobulges is obtained for most of the correlations when we added data from the literature since the value of the slopes dramatically varies in some cases by a factor of $\sim3$, as in the case of the bulge size and bulge luminosity, or even more significantly, as it occurs for the total luminosity correlation, where not only for pseudobulges subsample but also for the entire sample, a substantial change is appreciated. This behaviour of having a flatter slope is shown in the works of \cite{Saglia_et_al_2016, deNicola_et_al_2019}, considering larger samples. Nevertheless, in the $M_{\bullet}\texttt{-}V_{rot}$ correlation, the slopes are less affected if additionnal pseudobulges are added, with the caveat that it has a larger scatter than $M_{\bullet}\texttt{-}\sigma$ and $M_{\bullet}\texttt{-}L$, although comparable with the one of $M_{\bullet}\texttt{-}L_{tot}$.

Therefore, we highlight that the differences found in the correlations between classical bulges and pseudobulges (as established in our results from Figs. \ref{M-Lum} to \ref{M-Disc} and Table \ref{Table_results_fits_M-rels}), arise from the distinct evolutionary pathways of such objects, i.e., a merger-driven scenario for classical bulges as elliptical galaxies. In contrast, pseudobulges follow a secular evolution \citep[e.g.,][]{Kormendy-Ho_2013}.  Nevertheless, we caution that larger and homogeneous samples of pseudobulges are needed to test whether they follow the SRs for SMBH in more detail.

\begin{table*}
\caption{Candidates for Dynamical Measurements of BH Masses, ordered by increasing projected radius of influence.} 
\label{Table_candidates}
\begin{center}
\tabcolsep 1.10 pt
%\fontsize{7}{7}
\footnotesize
\begin{tabular}{@{\extracolsep{1.7pt}}l c c c c c c c c c }
%\small
%\begin{tabular}{@{\extracolsep{6pt}}l c c c c c c c c c }
\hline
{} & \multicolumn{3}{c}{$M_{\bullet}-L$} & \multicolumn{3}{c}{$M_{\bullet}-\sigma$} & \multicolumn{2}{c}{$M_{\bullet}-r_e$} & {} \\
\cmidrule{2-4}
\cmidrule{5-7}
\cmidrule{8-9}
Name & $M_{\bullet}({KH})$ & $M_{\bullet}({dN})$ & $M_{\bullet}$ & $M_{\bullet}({KH})$ & $M_{\bullet}({dN})$ & $M_{\bullet}$ & $M_{\bullet}({dN})$ & $M_{\bullet}$ & $~r_f$ \\
~~(1) & ~~~~(2) & ~~~~(3) & (4) & ~~~~(5) & ~~~~(6) & (7) & ~~~~(8) & (9) & (10) \\
\hline
UGC 2450 & 10.33 $\pm$ 0.02 & 10.04 $\pm$ 0.02 & 10.45 $\pm$ 0.02 & 09.65 $\pm$ 0.18 & 09.69 $\pm$ 0.21 & 10.02 $\pm$ 0.23 & 10.59 $\pm$ 0.03 & 10.04 $\pm$ 0.04 & 645 ($0\farcs40$) \\
Maffei1 & 07.72 $\pm$ 0.08 & 07.65 $\pm$ 0.07 & 07.62 $\pm$ 0.09 & 08.36 $\pm$ 0.08 & 08.20 $\pm$ 0.09 & 08.40 $\pm$ 0.09 & 08.44 $\pm$ 0.06 & 07.69 $\pm$ 0.06 & 6 ($0\farcs43$)\\
IC 5328 & 09.01 $\pm$ 0.06 & 08.83 $\pm$ 0.05 & 09.02 $\pm$ 0.06 & 08.55 $\pm$ 0.07 & 08.42 $\pm$ 0.08 & 08.64 $\pm$ 0.09 & 09.41 $\pm$ 0.09 & 08.75 $\pm$ 0.10 & 92 ($0\farcs50$) \\
NGC~2985 & 08.54 $\pm$ 0.02 & 08.40 $\pm$ 0.02 & 08.51 $\pm$ 0.02 & 07.82 $\pm$ 0.06 & 07.58 $\pm$ 0.07 & 07.73 $\pm$ 0.08 & 09.15 $\pm$ 0.02 & 08.47 $\pm$ 0.03 & 66 ($0\farcs66$) \\
IC 1101 & 10.62 $\pm$ 0.02 & 10.30 $\pm$ 0.02 & 10.77 $\pm$ 0.03 & 09.61 $\pm$ 0.06 & 09.64 $\pm$ 0.07 & 09.97 $\pm$ 0.08 & 10.63 $\pm$ 0.03 & 10.08 $\pm$ 0.04 & 1336 ($0\farcs76$) \\
NGC~2768 & 08.89 $\pm$ 0.03 & 08.72 $\pm$ 0.03 & 08.90 $\pm$ 0.04 & 08.31 $\pm$ 0.04 & 08.14 $\pm$ 0.04 & 08.34 $\pm$ 0.05 & 09.60 $\pm$ 0.01 & 08.95 $\pm$ 0.01 & 90 ($0\farcs91$) \\
UGC 2438 & 10.27 $\pm$ 0.02 & 09.99 $\pm$ 0.02 & 10.39 $\pm$ 0.03 & 08.76 $\pm$ 0.25 & 08.66 $\pm$ 0.29 & 08.90 $\pm$ 0.31 & 10.54 $\pm$ 0.03 & 09.99 $\pm$ 0.03 & 1438 ($0\farcs92$) \\
NGC 4874 & 09.96 $\pm$ 0.01 & 09.70 $\pm$ 0.01 & 10.05 $\pm$ 0.01 & 09.08 $\pm$ 0.03 & 09.03 $\pm$ 0.03 & 09.30 $\pm$ 0.04 & 10.29 $\pm$ 0.02 & 09.71 $\pm$ 0.02 & 491 ($0\farcs98$)\\
NGC~4365 & 09.31 $\pm$ 0.03 & 09.11 $\pm$ 0.03 & 09.35 $\pm$ 0.04 & 08.96 $\pm$ 0.02 & 08.90 $\pm$ 0.03 & 09.16 $\pm$ 0.03 & 09.71 $\pm$ 0.01 & 09.07 $\pm$ 0.01 & 122 ($1\farcs16$)\\
NGC~5084 & 09.41 $\pm$ 0.03 & 09.19 $\pm$ 0.03 & 09.45 $\pm$ 0.04 & 08.50 $\pm$ 0.05 & 08.36 $\pm$ 0.06 & 08.58 $\pm$ 0.07 & 09.55 $\pm$ 0.01 & 08.90 $\pm$ 0.01 & 246 ($1\farcs23$)\\
UGC 5515 & $10.45\pm 0.06$ & $10.15 \pm 0.06$ & $10.59 \pm 0.07$ & $09.26 \pm 0.14$ & 09.24 $\pm$ 0.17 & 09.53 $\pm$ 0.18 & 10.82 $\pm$ 0.11 & 10.29 $\pm$ 0.12 & 1297 ($1\farcs31$) \\
NGC~1291 & 08.66 $\pm$ 0.01 & 08.50 $\pm$ 0.01 & 08.63 $\pm$ 0.01 & 08.09 $\pm$ 0.21 & 07.89 $\pm$ 0.24 & 08.06 $\pm$ 0.26 & 09.53 $\pm$ 0.01 & 08.88 $\pm$ 0.02 & 66 ($1\farcs50$) \\
NGC~3551 & 10.23 $\pm$ 0.03 & 09.96 $\pm$ 0.03 & 10.36 $\pm$ 0.04 & 08.95 $\pm$ 0.13 & 08.88 $\pm$ 0.15 & 09.15 $\pm$ 0.16 & 10.98 $\pm$ 0.07 & 10.47 $\pm$ 0.08 & 1087 ($1\farcs56$) \\
NGC~4125 & 09.52 $\pm$ 0.03 & 09.30 $\pm$ 0.03 & 09.58 $\pm$ 0.04 & 08.73 $\pm$ 0.06 & 08.63 $\pm$ 0.07 & 08.87 $\pm$ 0.08 & 10.31 $\pm$ 0.03 & 09.73 $\pm$ 0.03 & 254 ($2\farcs30$) \\
MCG-02-12-039 & 10.38 $\pm$ 0.05 & 10.09 $\pm$ 0.05 & 10.51 $\pm$ 0.05 & 08.85 $\pm$ 0.11 & 08.76 $\pm$ 0.13 & 09.01 $\pm$ 0.14 & 10.99 $\pm$ 0.09 & 10.48 $\pm$ 0.09 & 1702 ($2\farcs43$) \\ \hline
\multicolumn{10}{l}{Columns: (1) Name of the galaxy. (2), (3), $\&$ (4) are BH masses estimates using $M_{\bullet}-L$ relation; (5), (6), $\&$ (7) BH masses }\\
\multicolumn{10}{l}{estimates using $M_{\bullet}-\sigma$ relation; (8) $\&$ (9) BH masses estimates using $M_{\bullet}-r_e$ relation; BH masses estimates using SRs from}\\
\multicolumn{10}{l}{\textit{KH}: \citet{Kormendy-Ho_2013}, \textit{dN}:\citet{deNicola_et_al_2019} and columns (4), (7) and (9) are from this work presented in Table \ref{Table_results_fits_M-rels};}\\
\multicolumn{10}{l}{(10) BH radius of influence ($r_f = GM_{\bullet}/\sigma^2$), where $M_{\bullet}$ is the mean of mass estimates per row; $r_f$ is given in pc and $r_f$ projected}\\
\multicolumn{10}{l}{ on the sky in seconds of arc shown in parentheses; the tabulated BH masses are given in $\logd[\mathrm{M_{\sun}}]$.}\\
%\hline
\end{tabular}
\end{center}
\end{table*}

\subsection{Interesting targets for follow-up observations}

Finally, we estimated SMBH masses using the SR listed in Table \ref{Table_candidates}, which involve structural parameters such as luminosity, velocity dispersion and effective radius. We used these relations because they have less scatter than others in Table \ref{Table_results_fits_M-rels}.

The goal is to carry out follow-up observations to identify attractive candidates that might host SMBHs with masses at least $M_{\bullet}\sim10^9\, \mathrm{M}_{\sun}$ through stellar- and gas-dynamical measurements at scales of the BH radius of influence, $r_f$, where the SMBH has a substantial impact on stellar and gas dynamics. To perform these kinds of observations is essential because SMBHs with such masses are not so common. For instance, from the recent compilation presented by \cite{deNicola_et_al_2019}, less than a third of their sample (comprising 83 SMBH detections) is above that threshold. Besides, these kinds of observations may provide critical tests for the applicability of SMBH scaling laws since many studies have reported SMBHs that are more or less massive than the predictions given by the scaling laws \citep{Lauer_et_al_2007, McConnell_et_al_2012, Lopez-Cruz_et_al_2014, Mehrgan_et_al_2019}.

Hence, from the sources listed in Table \ref{Table_candidates}, for NGC~2768, NGC~4365, NGC~5084, NGC~1291 and NGC~4125, the three scaling laws predict $M_{\bullet} \sim 10^9\; \mathrm{M}_{\sun}$ with a $r_f \gtrsim 1\farcs0$. On the other hand, IC~5328, NGC~2985, and Maffei1 have smaller radii and SMBH masses $M_{\bullet} \sim 10^8\,\mathrm{M_{\sun}}$; however, they are still interesting candidates for mapping the velocity field around the SMBH, as $r_f$ $\sim 0\farcs5$.

For the subsample of cD galaxies, we highlight the targets UGC~2438, NGC~4874, UGC~5515, NGC~3551 and MCG-02-12-039 also with a remarkable $r_f\gtrsim 1^{\prime\prime}$, for which Adaptive Optics (AO) assisted high-resolution observations with 8-10m ground-based telescopes can provide sufficient resolution and sensitivity to resolve their $r_f$. At the same time, the UGC~2450 and IC~1101 are also attractive targets with $r_f\gtrsim0\farcs5$. Regarding the BH masses of such cD galaxies, we estimated masses of $M_{\odot}$ $\sim 10^{10} M_{\odot}$, and in some cases particular cases, $M_{\bullet}$ $\sim 10^{11} M_{\odot}$ (NGC~3551 and MCG-02-12-039) according to the BH mass-effective radius relation.

 The mean values of the BH masses reported in Table \ref{Table_candidates} are used to generate $r_f$; however, the $M_{\bullet}\texttt{-}r_e$ relation tends to predict higher SMBH masses (close to an order of magnitude in some cases) than the other two relations. Thus, considering only SMBH masses in columns 7 to 9, the $r_f$ would be even larger. A similar effect when estimating SMBH masses using the size parameter can be appreciated in \cite{Lopez-Cruz_et_al_2014} since the size of the core of the cD galaxy Holm~15A (included in our sample) provides a $M_{\bullet} \sim 10^{11}\,\mathrm{M_{\sun}}$. In comparison, we estimated masses of $\sim 10^{9}$ and $\sim10^{10}\,\mathrm{M_{\sun}}$ with $\sigma$ and luminosity, only with the effective radius used as a proxy we get masses of $\sim 10^{11}$ $\mathrm{M_{\sun}}$, nevertheless, the $M_{\bullet}\texttt{-}r_e$ relation has a larger scatter than $M_{\bullet}\texttt{-}\sigma$ and $M_{\bullet}\texttt{-}L$ relations. However, \citeauthor{Lopez-Cruz_et_al_2014} cautiously suggested $M_{\bullet} \sim 10^{10}\,\mathrm{M_{\sun}}$ for Holm~15A because the SRs had not been extended to include supergiant galaxies as we do in this work. Indeed, \citet{Mehrgan_et_al_2019}'s study confirmed that Holm~15A does host an UMBH whose mass is $M_{\bullet} = (4.0 \pm 0.80) \times 10^{10}\; \mathrm{M}_{\sun}$. In fact, \citet{Lopez-Cruz_et_al_2014} suggested that cD galaxies might follow different SRs, which needs to be studied in more detail. Nevertheless, our cautionary suggestion from this work and \citeauthor{Lopez-Cruz_et_al_2014} is that SLABs should remain hypothetical.

\section{Summary and Conclusions}\label{Con}

In this work, we studied a sample of large galaxies observed in the NIR bands of 2MASS. The selection of galaxies covers most types, including supergiant galaxies. Only $K_s$ band measurements of cD galaxies have been included. We modelled the SB of galaxies through a two-dimensional photometric decomposition to obtain their structural parameters. Our results are also presented in the context of bulge types, namely, classical and pseudobulges. Details about 2D decomposition and the separation of bulge types are outlined in \citetalias{Rios-Lopez_et_al_2021}. In this paper, we reanalysed well-known galaxy SR and relations between the SMBHs and some global properties of galaxies. We present our main results below: 

\begin{enumerate}

\item We have demonstrated that 2MASS photometry is deep enough to study nearby giant galaxies, allowing the analysis of subcomponents such as bulges, discs and bars. We have considered previous problems with LGA mosaics and compared our analysis with other studies that included deeper NIR observations.

\item  We have revisited SR for a sample of galaxies observed in the NIR bands: Fundamental Plane (FP)  projections, such as the KR, FJR and the FP itself, and the LCR, LSR, and CMR, among others. We highlight in our plots between classical bulges and pseudobulges according to our classification presented in \citetalias{Rios-Lopez_et_al_2021} to recognise the trend displayed by these two populations of bulges. We found that our correlations are consistent with previous works. 

\item Our results also confirm the behaviour of classical bulges since these follow FP scaling relations traced by elliptical galaxies. At the same time, most pseudobulges are seen as outliers in the low-regime of such correlations for parameters such as luminosity or velocity dispersion, among others. In general, we found that the FP does not distinguish between bulge populations, while its projections, mainly KR, LSR, FJR, and CMR, are more suitable for differentiating bulge types. 

\item The study of the CMR has allowed us to introduce a new criterion for bulge classification. We found the NIR colours of classical bulges and pseudobulges are constant:  $\langle{J-H}\rangle=0.7$, $\langle H-K_s \rangle=0.3$, and $\langle J-K_s \rangle=1.0$; nevertheless, classical bulges are more luminous than $M_{K_s}=-22 \; \mathrm{mag}$ and pseudobulges are less luminous than this threshold.

\item Using the SR from this work, we suggest that the mass of the disc is $M_{dics}\propto L^{\frac{4}{3}}$; then, using the CMR for discs, we found that more massive discs are redder \citep[cf.,][]{2001ApJ...550..212B}. We suggest that regardless of the aperture size, colour apertures will include the disc component. Hence, previous studies of the CMR in the NIR missed that bulges from ETGs and LTGs have the same colours in a range of about nine magnitudes. Thus, the main contribution to the CMR's slope for LTG and ETG galaxies comes from the disc component, where the history of stellar formation proceeds inside-out \citep[e.g.,][]{2014A&A...562A..47G}. In the optical, the CMR for E+S0s has been regarded as a metallicity effect \citep[e.g.,][]{1997A&A...320...41K, Lopez-Cruz_et_al_2004}. However, we found, for the first time, that bulges have a flat CMR in the NIR bands. This is expected if NIR colours are insensitive to the metallicity effects.

\item We found that the slopes of the luminosity-size relations presented in this work are very close considering the entire sample ($\sim$1 for the LSR, while for the LSR$_d$ is $\sim$1.3), a result similar for classical bulges ($\sim$0.95). However, the slope of the LSR for pseudobulges is $\sim$0.20, which might indicate a hint on the evolution pathways for pseudobulges in the sense that pseudobulges do not simply have similar physical properties as galaxy discs. Still, they represent a complex transient morpholgical feature similar to galactic bars, oval discs, inner and outer rings, among others \citep[e.g.,][]{Bournaud_et_al_2005, 2013seg..book....1K}.

\item Advanced hydrodynamical numerical simulations show that ex situ processes dominate the growth ($\sim$70 \%) of cD galaxies. A detailed inspection of the SRs shows that cD galaxies are not simple extensions of classical bulges, in agreement with previous studies \citep[e.g.,][]{1987ApJS...64..643S}. Therefore, we conclude that our observations support \cite{2023MNRAS.521..800M} model.

\item We also studied SR for SMBHs, such as $M_{\bullet}$ with luminosity, effective radius, S\'ersic index, velocity dispersion, and rotation velocity. We report that our findings are in agreement with previous works. Furthermore, pseudobulges populate the lower part of the correlations hosting less massive BHs than classical bulges, which is in agreement with previous studies; this reinforces our bulge classification (see \citetalias{Rios-Lopez_et_al_2021}).

\item We found that the $M_{\bullet}\texttt{-}\sigma_e$ has the lowest intrinsic scatter among the SMBH SR presented here. This result agrees with previous works \citep[e.g.,][]{Ferrarese-Merritt_2000, Gebhardt_et_al_2000, Kormendy-Ho_2013, Saglia_et_al_2016, deNicola_et_al_2019}. Regarding the $M_{\bullet}$-$L$ relation for the bulge, after $M_{\bullet}\texttt{-}\sigma$, it has a smaller scatter than the other correlations presented here, which also coincides with previous studies \citep[e.g.,][]{Marconi-Hunt_2003, Kormendy-Ho_2013, deNicola_et_al_2019}. Then, we conclude that the $M_{\bullet}\texttt{-}\sigma_e$ relation is the most accurate SMBH mass estimator.

\item Besides, we show that the total luminosity of the galaxy and the bulge size  (effective radius) correlate with the $M_{\bullet}$, although with a larger scatter compared to $\sigma$ and luminosity of bulge.

\item The relation between $M_{\bullet}$ and $n$ has a high dispersion; this was also found in previous studies \citet[e.g.,][]{Beifiori_et_al_2012, Kormendy-Ho_2013}. The evident high dispersion in this correlation can be explained by considering that parameter coupling introduces inaccuracies in determining the Sérsic index $n$ in the fits to the galaxies' SB, in general. Hence, the lack of correlation in the $M_{\bullet}\texttt{-}n$ diagram was unsurprising. 

\item In addition, we explored the relations of SMBH masses with disc parameters, such as $V_{rot}$ and $L_{disc}$, and we found poor correlations, similar to previous results \citep[][]{Beifiori_et_al_2012, Kormendy-Ho_2013}. Our findings also confirm that SMBH masses do not correlate with the disc of galaxies, or if they do, to a lesser extent. However, a more detailed study may be necessary for a more robust conclusion. 

\item We highlight the differences in the correlations between classical bulges and pseudobulges, as established in our results from Figs. \ref{M-Lum} to \ref{M-Disc} and Table \ref{Table_results_fits_M-rels}, arise from the distinct evolutionary pathways of bulge types. While pseudobulges show a trend which is poorly consistent with BH scaling relations, nevertheless, we recall the fact of having a small subsample of eight pseudobulges; then, to explore this trend, we added more pseudobulges with data from the literature, and our results indicate still different correlations for pseudobulges. Therefore, testing this result in more detail with a larger sample of pseudobulges and observations that resolve the nuclear dynamics becomes relevant.

\item Regarding cD galaxies, these galaxies appear as extremes on the SR considered here. Only three of the 19 cD galaxies in this study have dynamically measured BH masses, namely Holm~15A, NGC~4889 and M~87, which host the most massive BHs in our sample. We suggest that cDs may not follow the same  BH's SR for the galaxy's total luminosity (Fig. \ref{M-Lum_tot}) and possibly for the velocity dispersion (Fig. \ref{M-sig}), following earlier suggestions  \citep[e.g.,][]{Lopez-Cruz_et_al_2014, Hlavacek-Larrondo_et_al_2012}. Nevertheless, we cannot confirm this as more dynamical masses for BH in cD galaxies are needed; Table \ref{Table_candidates} provides a few suitable candidates (see below). 

\item Finally, we singled out candidates suitable for dynamical measurements of BH masses using stellar and gas tracers. These targets are presented in Table \ref{Table_candidates} with their corresponding SMBH masses estimated using SRs from this work and the literature. Expected values for the BH radius of influence are also given, which, with A0-assisted 8 to 10 m telescopes equipped with moderate resolution IFU, can be spatially resolved to obtain high-quality spectra at those angular scales. The large photon ring radii (${\sqrt{27} GM_{\bullet}}/{c^2}$) of galaxies Holm~15A, IC~1101 and MCG-02-12-039 make them attractive targets for the Event Horizon Telescope if the angular resolution is improved.

\item The results presented in this paper make sense in the light of the framework described in depth by \citet{2013seg..book....1K}. Hence, bulges, pseudobulges, discs and supergiant galaxies, as hosts of SMBHs, conform to different populations whose evolution can at least be envisaged with firmly established SRs.

\end{enumerate}

\section{Acknowledgements}

%We are grateful to the anonymous Referee's careful review and insight, which helped us clarify and significantly improve this paper. 

We are grateful to the anonymous Referee's careful review and insight, which helped us clarify and significantly improve this paper. Our most profound appreciation goes to the Referee for suggesting that we, too, dedicate this paper to Prof. Tom Jarrett.

ERL was supported by a PhD research grant from the Mexican Research Council (CONACyT) and also acknowledges Dr. Elena Terlevich for the SNI-CONACyT graduate assistantship. MV acknowledges support from the CONAHCYT grant from the program ``Estancias Posdoctorales por M\'exico 2021''. CA research is funded by Universidad Autónoma de Sinaloa through project PROFAPI 2022, with the project key A1009. The work of VRC and EARH was supported by {\em Galaxy Maquila}, an undergraduate Summer Research Program at INAOE led by OLC. We thank Dr X. Hern\'andez-Doring for constructive discussions and suggestions on many issues of this paper. ERL, CA, MV, and OLC acknowledge support from Mexico's Sistema Nacional de Investigadoras e Investigadores (SNII). 

This publication uses data products from the Two Micron All Sky Survey, a joint project of the University of Massachusetts and the Infrared Processing and Analysis Center/California Institute of Technology, funded by the National Aeronautics and Space Administration and the National Science Foundation. We also acknowledge using the NASA/IPAC Extragalactic Database (NED) and \texttt{HyperLeda} databases.

We humbly dedicate this paper to Prof. John Kormendy for his outstanding contributions to further our knowledge of galaxy evolution. We recognise his engaging, exemplary research approach, which was distinguished through careful measurements of galaxies and their structures and keen interpretation of the results. His utter dedication has been a constant source of inspiration and encouragement throughout our careers. 

With great sadness, we also wish to dedicate this paper to Prof. Thomas Harold Jarrett, who suddenly passed away on July 1st, 2024. Our academic and research community deeply feels his loss, as he was a highly accomplished researcher and the mastermind behind the 2MASS Large Galaxy Atlas, which was fundamental to this series of papers. Tom was our longtime friend and esteemed colleague, always willing to help and share his knowledge up to the last hours of his life. We will always miss him.

\section{Data Availability}
The data used in this paper are available on tables and online supplementary material. Supplement materials were gathered from sources in the public domain: 2MASS Large Galaxy Atlas at \url{https://irsa.ipac.caltech.edu/applications/2MASS/LGA/}, 2MASS image tiles at \url{https://irsa.ipac.caltech.edu/applications/2MASS/IM/interactive.html#pos}, NASA Extragalactic Database at \url{http://ned.ipac.caltech.edu/} and \texttt{HyperLeda} database at \url{http://leda.univ-lyon1.fr/leda/param/vdis.html}.

\clearpage

\bibliography{Paper-II.bib}
%\label{lastpage}

%\appendix\label{Append}

\end{document}